\documentclass[11pt,a4paper]{article}
\usepackage{jcappub}

\usepackage{booktabs}
\usepackage{amsmath}
\usepackage{amssymb}
\usepackage{hhline}  
\usepackage[scale={.75,.75}]{geometry}
\usepackage{url}
\usepackage{caption} 
\usepackage[numbers, sort&compress]{natbib}
\usepackage{epsfig}
\usepackage{lscape}
\usepackage{mathrsfs}
\usepackage{multirow}
\usepackage{physics}
\usepackage{color}
\usepackage{verbatim}
\usepackage{nicefrac} 
\usepackage{upgreek}
\usepackage{hyperref}
\usepackage{enumitem}
\usepackage{outlines}
\usepackage{slashbox,booktabs,amsmath}
\usepackage{multirow}

\usepackage{makecell}

\newcommand{\varphio}{\bar{\varphi}}

\newcommand{\q}{\textbf{q}}

\newcommand{\p}{\textbf{p}}

\newcommand{\x}{\textbf{x}}

\newcommand{\V}{\mathcal{V}}

\allowdisplaybreaks
\usepackage{tikz}
\usetikzlibrary{patterns}
\usetikzlibrary{calc,arrows,positioning}
\usetikzlibrary{arrows,shadows}
\usepackage{circuitikz} 
\usetikzlibrary{decorations.pathmorphing}	
\usetikzlibrary{decorations.markings}
 
\tikzset{
	 >=latex, 
    vector/.style={decorate, decoration={snake,amplitude=2pt,segment length=1.7mm}, draw},
	provector/.style={decorate, decoration={snake,amplitude=2.5pt}, draw},
	antivector/.style={decorate, decoration={snake,amplitude=-2.5pt}, draw},
    fermion/.style={draw=black, postaction={decorate},
        decoration={markings,mark=at position .55 with {\arrow[draw=black]{>}}}},
    fermionbar/.style={draw=black, postaction={decorate},
        decoration={markings,mark=at position .55 with {\arrow[draw=black]{<}}}},
    fermionnoarrow/.style={draw=black},
    gluon/.style={decorate, draw=black,
        decoration={coil,amplitude=4pt, segment length=5pt}},
    scalar/.style={dashed,draw=black, postaction={decorate},
        decoration={markings,mark=at position .55 with {\arrow[draw=black]{>}}}},
    scalarbar/.style={dashed,draw=black, postaction={decorate},
        decoration={markings,mark=at position .55 with {\arrow[draw=black]{<}}}},
    scalarnoarrow/.style={dashed,draw=black},
    electron/.style={draw=black, postaction={decorate},
        decoration={markings,mark=at position .55 with {\arrow[draw=black]{>}}}},
	bigvector/.style={decorate, decoration={snake,amplitude=4pt}, draw},
}

\title{ 
General Markovian Equation for Scalar Fields in a Slowly Evolving Background
}

\author[a]{Gilles Buldgen}
\author[a]{Marco Drewes}
\author[b,c]{Jin U Kang}
\author[b]{Ui Ri Mun}

\affiliation[a]{Centre for Cosmology, Particle Physics and Phenomenology,
Universit\'e catholique de Louvain, Louvain-la-Neuve B-1348, Belgium
}
\affiliation[b]{Department of Physics, Kim Il Sung University,
Ryongnam Dong, TaeSong District, Pyongyang, DPR Korea
}
\affiliation[c]{
International Centre for Theoretical Physics,
 Strada Costiera 11, 34151 Trieste, Italy
}

 \date{}

\abstract{
We present a general and model-independent method to obtain an effective Markovian quantum kinetic equation for the expectation value of a slowly evolving scalar field in an adiabatically evolving background from first principles of nonequilibrium quantum field theory. The method requires almost no assumptions about the field's interactions and the composition of the background, except that 1) the coupling constants shall be small enough for perturbation theory to be applicable, 2) there is a clear separation between microphysical time scales and the rate at which bulk properties change, and 3) higher time derivatives of the field remain small. The resulting Markovian equation of motion is expressed in terms of an effective potential and friction coefficients. Motivated by cosmological applications we focus on spatially homogeneous and isotropic systems, but the approach could also be applied to spatial gradients.}    
           
\begin{document}      
 \maketitle

\section{Introduction}\label{introduction}

Scalar and pseudoscalar fields play an important role in many areas of science. Prominent examples include the Landau theory of phase transitions \cite{Landau:1937obd}, the Ising model \cite{Lenz:460663,Ising:1925em}, the Landau-Ginzburg theory of superconductivity \cite{Ginzburg:1950sr}
or the Brout-Englert-Higgs mechanism~\cite{Englert:1964et, Guralnik:1964eu, Higgs:1964pj}.  With a Wilsonian approach to renormalisation \cite{Wilson:1973jj}, they can be used to effectively describe many macroscopic properties of quantum systems that are controlled by order parameters \cite{ZinnJustin:2002ru}, such as superconductivity and magnetic properties. Scalar fields also play an important role in theories beyond the Standard Model (SM) of particle physics, and in particular in cosmology. For instance, they may explain the strong CP-problem \cite{Peccei:1977hh},  are candidates for Dark Matter  \cite{McDonald:1993ex, Burgess:2000yq, Bento:2000ah, Arias:2012az} or Dark Energy \cite{Wetterich:1987fm, ArmendarizPicon:2000dh, Kang:2007vs, Copeland:2006wr} and can drive cosmic inflation \cite{Starobinsky:1980te,Guth:1980zm,Linde:1981mu, ArmendarizPicon:1999rj}. The low-energy effective description of string theory and other theories involving extra dimensions generally include numerous scalar fields, called moduli, that parameterize the properties of the compactified internal dimensions (see e.g. \cite{Douglas:2015aga, Denef:2007pq} for a review).

It has been known for a long time that coupling to a time-dependent background leads to the production of particles \cite{SCHRODINGER1939899}. If the time evolution is non-adiabatic, the non-perturbative particle production is unavoidable and usually described by the Bogoljubov-Valatin transformation \cite{Bogoljubov1958,Valatin1958}. However, there are many situations in which the bulk properties of the system evolve slowly with respect to other relevant microscopic time scales, such as the frequencies of elementary excitations in the system under consideration. On general grounds one would expect that the evolution of the bulk properties can effectively be described by Markovian equations that include effective friction terms and effective potentials. Considerable effort has been made in different areas of science to derive  quantum kinetic equations of this kind, cf.~Refs.~\cite{Chou:1984es,Calzetta:2008iqa} for reviews.

Many of these works are concerned with the evolution of particle occupation numbers, cf.~e.g.~\cite{Calzetta:1986cq,Ivanov:1999tj,Arnold:2002zm,Juchem:2003bi,Berges:2005md,Herranen:2008hu,Lindner:2005kv,Garbrecht:2011xw,Drewes:2012qw}.  
The present work is focused on the evolution of mean fields
 in the context of relativistic quantum field theory, which has also been the subject of several studies in the past.
 The authors of  
 \cite{Morikawa:1986rp,Calzetta:1989vs,Boyanovsky:1994me,Greiner:1996dx} primarily focus on the regime where a scalar field $\phi$ is placed in a thermal bath and performs oscillatory motion. They obtain Markovian equations when the elongation is small enough to linearise the equation of motion, so
 that these oscillations are harmonic. This approach can be extended to the mildly non-linear regime \cite{Ai:2021gtg}.  
 In  \cite{Yokoyama:1998ju,Yokoyama:2004pf,BasteroGil:2010pb,Mukaida:2013xxa,Cheung:2015iqa} the case of a slowly evolving scalar field $\phi$ is studied using  similar methods as employed here, but the authors restrict themselves to specific models, perform a Taylor expansion around the ground state in field space, and assume that the background is in thermal equilibrium. 
 The present work generalises this approach in a model independent way. The method 
 requires  no assumption about the spin of the degrees of freedom that the field under consideration couples to, nor on the type   
of their interactions, and no quasiparticle approximation is needed.

We consider the expectation value $\varphi=\langle\phi\rangle$ of a  real scalar field $\phi$ to illustrate our approach. 
Quantum fluctuations of $\phi$ around the expectation value $\varphi$ are described by a field $\eta$ which we define as $\phi = \varphi + \eta$. 
This example has a number of direct applications in cosmology, such as the evolution of the inflaton field during \cite{Berera:1995ie,Bastero-Gil:2012akf} and after \cite{Kolb:2003ke,Drewes:2013iaa,Mukaida:2012bz,Harigaya:2014waa,Drewes:2014pfa,Drewes:2019rxn,Co:2020xaf,Garcia:2020wiy} cosmic inflation, the evolution of moduli fields in the post-inflationary epoch \cite{Buchmuller:2004xr,Buchmuller:2004tz,Bodeker:2006ij} or cosmic phase transitions \cite{Curtin:2016urg}. 
Motivated by these cosmological applications we restrict ourselves to homogeneous and isotropic systems and only consider the evolution in time, noting that spatial gradients could be treated in an analogous manner.
Beyond the domain of cosmology, the approach presented here may be applied to any system that can effectively be described by scalar fields, including systems in condensed matter or atomic physics. Explicit examples for systems described by scalar field theory that can exhibit both, adiabatic and non-adiabatic behaviour, include \cite{Chatrchyan:2020syc,Prufer:2019kak,Knap,Scammell_2017,Scammell:2015swa}.

Our main assumption is that $\varphi$ changes sufficiently slowly such that the expansion
\begin{equation}
\varphi(t')^n =  \varphi(t)^n + n (t'-t)\dot{\varphi}(t)\varphi(t)^{n-1}  
+ \mathcal{O}[\ddot{\varphi}]\,, \label{AD-approx}
\end{equation}
can be justified inside loop integrals. The physical interpretation of this assumption is that $\varphi(t)$ does not change much in the course of a single interaction. Mathematically it relies on the observation that $\varphi(t)$ in collision integrals is always convolved 
with other functions that can be expressed as products of correlators, and that those convolutions are suppressed for separations of the time arguments that are much larger than the typical microscopic time scale $\tau_{\rm int}$ in the system. They therefore act as ``window functions" in time that suppress contributions to memory integrals for large separations $\gg\tau_{\rm int}$ of the arguments, leading to equations that are effectively local in time.

In the following we outline a general method to derive an effective quantum kinetic equation of the form\footnote{\label{NoiseFootnote}
Based on the fluctuation-dissipation theorem one may expect noise terms on the right hand side of \eqref{EOM-slow-roll-LTE}. The reason why they do not appear is that the definition of $\varphi=\langle\phi\rangle \equiv {\rm Tr}(\varrho \phi)$
includes an average over statistical fluctuations included in the von Neumann density operator $\varrho$.
A noise term indeed appears in the equation of motion for non-averaged quantities, cf.~\cite{Chou:1984es,Calzetta:2008iqa}. For the specific example of the field $\phi$ it yields a Langevin type equation that has e.g.~been discussed in Refs.~\cite{Morikawa:1986rp,Greiner:1998vd,Yokoyama:2004pf,Boyanovsky:2004dj,Anisimov:2008dz,Gautier:2012vh}.
} 
\begin{equation}
\ddot{\varphi} 
+ \sum \Gamma_{\varphi}^{(n)}
\dot{\varphi}^n 
+ \partial_{\varphi} \mathcal{V}_\varphi 
= 0.   \label{EOM-slow-roll-LTE}
\end{equation}
Here $\mathcal{V}_\varphi$ is an effective potential for $\varphi$, and the $\Gamma_{\varphi}^{(n)}$ represent friction coefficients.

This article is organised as follows. 
In Sec.~\ref{2PI_GENERAL_SECTION} we review basic ingredients of the two particle irreducible (2PI) effective action formalism, roughly following Ref.~\cite{Berges:2004yj}. 
In Sec.~\ref{sec:markovian} we present our main result, a master formula for the derivation of a Markovian equation of motion for $\varphi$.
In Sec.~\ref{sec:LeadingTerm} we provide a diagrammatic interpretation of the leading friction coefficient.
In Sec.~\ref{sec:model} we explicitly compute this term in a specific model.
In Sec.~\ref{sec:Conclusion} we conclude.
In Appendix \ref{appendix} we discuss the range of validity of our approach in some more detail.

\section{2PI formalism for an interacting scalar}
\label{2PI_GENERAL_SECTION}

We consider a model with an unspecified number of real scalar fields $\Phi_a$. We make no assumptions on their interactions except that the coupling constants shall be small enough that perturbation theory can be applied. For simplicity we assume that only one of the $\Phi_a$ has a non-vanishing expectation value.
We identify that field with $\phi$ from  Sec.~\ref{introduction} and label the expectation value by $\varphi$. Lifting this assumption is straightforward, but complicates the equations considerably.
The generating functional reads
\begin{equation}
Z[J,R]=\int \mathcal{D}\Phi \exp\left[i\left(S[\Phi]+\int_x J_a(x)\Phi_a(x) + \frac{1}{2}\int_{xy}R_{ab}(x,y)\Phi_a(x)\Phi_b(y)\right)\right].
\end{equation}
Here $\Phi$ without index collectively refers to the set $\{\Phi_a\}$, and $J_a$ and $R_{ab}$ are sources.
$S[\Phi]$ is the classical action, for simplicity we assume that it has been brought in a form with canonical kinetic terms, $S[\Phi] = \frac12 \partial_\mu\Phi_a\partial^\mu\Phi_a - V(\Phi)$, where $V(\Phi)$ contains all interaction terms between the fields.
$\int_x=\int_{\mathcal{C}}\int d^3\mathbf{x}$ is a 4-dimensional spacetime integral, and $\mathcal{C}$ a time-path that is to be chosen to suit the problem of interest. We obtain the 2PI effective action from $W[J,R]=-i\ln Z[J,R]$ by performing a double Legendre transform of $W[J,R]$ with respect to the sources\footnote{Here the trace operator implicitly sums over all field degrees of freedom labels, and integrates spacetime arguments.}
\begin{equation}
\mathbf{\Gamma}[\varphi ,\Delta]= W[J,R] - \int_x \varphi_a(x)J_a(x) - \frac{1}{2}\int_{xy}\varphi_a(x)\varphi_b(y)R_{ab}(x,y) - \frac{1}{2}\text{Tr}[\Delta R].
\end{equation}
From the 2PI effective action we can obtain the equations of motion for $\varphi$ as well as all correlation functions
$\Delta_{ab}(x,y)=\langle{\rm T}_{\mathcal{C}}\Phi_a(x)\Phi_b(y)\rangle$ with time ordering along the contour $\mathcal{C}$. It is convenient to split $\mathbf{\Gamma}[\varphi ,\Delta]$ into a tree level part $S[\varphi]$, a one-loop correction $\mathbf{\Gamma}_1[\varphi,\Delta]$ and a piece that contains all terms with two or more loops $\mathbf{\Gamma}_2[\varphi, \Delta]$,
\begin{equation} \label{2PI-action}
{ \mathbf{\Gamma}[\varphi, \Delta] \equiv  S[\varphi]+ \mathbf{\Gamma}_{\rm loop }[\varphi,  \Delta] } =  S[\varphi]+ \mathbf{\Gamma}_1[\varphi, 
\Delta]+\mathbf{\Gamma}_2[\varphi, \Delta].
\end{equation}
Note here that only $ \mathbf{\Gamma}_{\rm loop}$ is a functional of $\Delta$. $S[\varphi]$ is sometimes referred to as the \emph{classical action}, though it really is the classical action functional with all fields $\Phi_a$ replaced by their quantum and statistical expectation values $\langle\Phi_a\rangle$. The one-loop term corresponds to the correction that leads to the Coleman-Weinberg potential \cite{Coleman:1973jx} and can be evaluated as
\begin{eqnarray}\label{Gamma1}
\mathbf{\Gamma}_1[\varphi,  \Delta]=\frac{i}{2} \text{Tr} \ln \left( \Delta^{-1}\right) +\frac{i}{2} \text{Tr} \left(G_0^{-1}[\varphi] \,  \Delta\right)\,
\end{eqnarray}
where $G_{0,ab}^{-1}[\varphi]$ are the inverse tree-level operators of the shifted action
\begin{equation}\label{tree_level_delta}
   i G_{0,ab}^{-1}[\varphi](x,y)\equiv \left.\frac{\delta^2 S[\Phi]}{\delta \Phi_a(x)\delta\Phi_b(y)}\right|_{\langle\Phi\rangle}.
\end{equation}
The subscript $_{\langle\Phi\rangle}$ indicates that all fields are to be evaluated at their expectation values. Note here that the operators $G_{0,ab}^{-1}[\varphi]$ are to be interpreted and dealt with as explicit $\varphi-$dependent contributions to the 2PI effective action. Partial functional derivatives with respect to $\varphi$ act on $G_{0,ab}^{-1}[\varphi]$. The $\Delta_{ab}(x,y)$, which represent the full (resummed) connected two-point functions, are formally regarded as independent dynamical variables with respect to the one-point functions in the 2PI formalism, as they stem from the $J$- and $R$- Legendre transforms independently.

The equations of motion for $\varphi$ 
 and the two-point functions can be obtained by functional differentiation of $\mathbf{\Gamma}[\varphi, \Delta]$.
In the present work we assume for simplicity that the external sources vanish, i.e., $J_a=0$ and $R_{ab}=0$,
so that
\begin{eqnarray}\label{EOM-varphi0}
0= \frac{\delta \mathbf{\Gamma}[\varphi, \Delta]}{\delta \varphi(x)}=\frac{\delta S[\varphi]}{\delta \varphi(x)} 
+ \frac{\delta \mathbf{\Gamma}_{\rm loop}[\varphi, \Delta]}{\delta \varphi(x)}= - \square \varphi(x) - \partial_\varphi V[\varphi(x)]+ \frac{\delta \mathbf{\Gamma}_{\rm loop}[\varphi, \Delta]}{\delta \varphi(x)}.
\end{eqnarray}
 and
\begin{eqnarray} \label{EOM-G}
\frac{\delta \mathbf{\Gamma}[\varphi,  \Delta]}{\delta  \Delta_{a b}(x,y)} =0 \,.
\end{eqnarray}
From Eq.~\eqref{2PI-action}, the equation of motion \eqref{EOM-G} becomes 
\begin{equation} \label{Inverse_G}
 \Delta^{-1}_{ab}(x, y)= G^{-1}_{0,ab}[\varphi](x,y) - \Pi_{ab}[\varphi, \Delta](x,y) \,,
\end{equation}
with the self-energies
\begin{equation}\label{PiDef}
\Pi_{ab}[\varphi,\Delta](x,y) = 2 i \frac{\delta \mathbf{\Gamma}_2[\varphi,\Delta]}{\delta \Delta_{ab}(x,y)}.
\end{equation}
By convolving Eq.~\eqref{Inverse_G} with $ \Delta_{bc}(z,y)$ (i.e. integrating over $z$ and summing over $b$) and using
$\sum_{b}\int_z  \Delta^{-1}_{ab}(x,z)\Delta_{bc}(z,y) = \delta_{ac}\delta_\mathcal{C}(x-y)$, one obtains a Schwinger-Dyson equation 
\begin{equation}\label{EOMofG}
\sum_b \int _z G^{-1}_{0,ab}[\varphi](x,z) \Delta_{bc}(z, y)
- \sum_{b}\int_z \Pi_{ab}(x,z;\varphi,\Delta)  \Delta_{bc}(z,y) 
=  \delta_{ac} \delta_\mathcal{C}(x-y) \,.
\end{equation}
Here $\delta_\mathcal{C}(x-y)$ is the four-dimensional delta function with time arguments on the contour $\mathcal{C}$. In the 2PI formalism one- and two-point functions are a priori independent quantities.
For instance, for a system with only one scalar field, $\mathbf{\Gamma}[\varphi,  \Delta]$ is a functional of two independent functions $\varphi$ and $\Delta$, which obey the equations \eqref{EOM-varphi0} and \eqref{EOMofG}. Practically these equations are solved perturbatively by expressing $\mathbf{\Gamma}_{\rm loop}$ and $\Pi$ in terms of Feynman diagrams, i.e., integrals over products made of $\Delta$ and $\varphi$. The equations of motion \eqref{EOM-varphi0} and \eqref{EOMofG}
then form a set of coupled integro-differential equations for the two functions $\Delta$ and $\varphi$.
The solution of Eq.~\eqref{EOMofG} at any given order in perturbation theory can formally be expressed as a functional of $\varphi$. This solution $\Delta[\varphi]$ is the fully resummed propagator in the presence of a background field $\varphi$.\footnote{
In general $\varphi$ and $\Delta$ can also depend on external sources, which we have set to zero to obtain \eqref{EOM-varphi0} and \eqref{EOM-G}, cf.~e.g.~\cite{Garbrecht:2015cla,Millington:2019nkw} for a discussion.} 
When plugging $\Delta[\varphi]$ back into \eqref{EOM-varphi0}, we obtain a generating functional $\mathbf{\Gamma}[\varphi,\Delta[\varphi]]$ for $\varphi$ at the desired order in perturbation theory.\footnote{In contrast to $\mathbf{\Gamma}[\varphi, \Delta]$
in \eqref{EOMofG}, $\mathbf{\Gamma}[\varphi,\Delta[\varphi]]$ cannot be used as a generating functional for $\Delta$ anymore since $\Delta[\varphi]$ already is the solution of its 2PI equation of motion. 
}  
The implicit dependence of the solution $\Delta[\varphi]$ on $\varphi$ has no effect on the functional derivative in \eqref{EOM-varphi0},  
\begin{equation}
\frac{\delta \mathbf{\Gamma}[\varphi,\Delta[\varphi]]}{\delta \varphi(z)}=
\frac{\partial \mathbf{\Gamma}[\varphi,\Delta[\varphi]]}{\partial\varphi(z)}+ \int_{xy}\underbrace{
\frac{\partial \mathbf{\Gamma}[\varphi,\Delta]}{\partial \Delta_{ab}(x,y)}\Bigg|_{\Delta[\varphi]}
}_{=0 \text{ by  virtue  of  Eq.\eqref{EOM-G}}}\frac{\delta\Delta_{ab}[\varphi](x,y)}{\delta \varphi(z)}=
\frac{\delta \mathbf{\Gamma}[\varphi,\Delta]}{\delta \varphi(z)}\Bigg|_{\Delta[\varphi]}
\end{equation}
and the resulting equation of motion remains the same. The same conclusion can be drawn if we had solved for $\varphi$ first.

\section{Markovian equation for a slowly evolving scalar}\label{sec:markovian}
We now proceed to derive 
a Markovian equation of motion for $\varphi$ from $\mathbf{\Gamma}[\varphi,\Delta[\varphi]]$.
This does not require explicit knowledge of the solution $\Delta[\varphi]$, we only need to use the fact that the resummed propagator can formally be expressed as a functional of $\varphi$, and that $\varphi$ changes slow enough that the gradient expansion \eqref{AD-approx} can be applied \emph{inside loop integrals}. We start with a functional Taylor expansion of $\mathbf{\Gamma}[\varphi,\Delta[\varphi]]$ 
around
a point $\bar{\varphi}$ in functional space, such that $\varphi = \bar{\varphi} + \delta\bar{\varphi}$ and
\begin{align}\label{FunctionalTaylor}
   \left.\frac{\partial \mathbf{\Gamma}_{\rm loop}[\varphi,  \Delta[\varphi]]}{\partial \varphi(x)}\right|_{
\substack{\varphio +  \delta\bar{\varphi}} }
& = \left. \frac{\partial \mathbf{\Gamma}_{\rm loop}[\varphi,  \Delta[\varphi]]}{\partial \varphi(x)} \right|_{\varphio}\nonumber \\
  + \sum_{n=1}^\infty \frac{1}{n!}
 \prod_{i=1}^n & \left[\int_{\mathcal{C}}  dx_i^0  
 \int d^3 \x_i \; \delta\bar{\varphi}(x_i)\right]\left[
 \frac{\delta^n}{\delta \varphi(x_1) \cdots \delta\varphi(x_n)}\left(\left.\frac{\partial\mathbf{\Gamma}_{\rm loop}[\varphi,  \Delta[\varphi]]}{\partial \varphi(x)}
 \right)\right]\right|_{\bar{\varphi}}.
\end{align}
Here the partial functional derivatives $\partial / \partial \varphi$ only apply to the explicit dependence of $\mathbf{\Gamma}_{\rm loop}$, not to the implicit dependence through $\Delta[\varphi]$. Diagrammatically the series \eqref{FunctionalTaylor} corresponds to expanding the $\varphi$-dependent vertices and propagators in loop integrals around their values at some reference point $\varphio$. This can be used as a perturbative approximation if a truncation at finite order can be justified. This is always the case if one is only interested in small deviations $\delta\bar{\varphi}$ from $\varphio$. Such a \emph{small field expansion} is widely used in the literature. The disadvantage is that the validity of this approximation is, by definition, only justified for a limited range of field excursions. There are, however, many situations in which one wants to track the evolution of the field over macroscopic times, such as cosmic inflation.

In the present work we take an alternative approach that utilises the fact that the functions with which $\varphi$ is convolved
in the memory integrals are suppressed for separations of the time arguments that are larger than some characteristic time $\tau_{\rm int}$. 
For this purpose we identify $\varphio$ with the constant function that takes the value of the self-consistent solution of the equation of motion evaluated at the reference time $t$, $\varphio=\varphi(t)$. Here $t\equiv x^0$ set by the local time argument of the equation of motion.
Usually $\tau_{\rm int}$ is closely related to the microscopic time scales in the system. 
The precise value of $\tau_{\rm int}$ and the functional form of the suppression are model dependent, cf.~appendix \ref{appendix}. A conservative choice is to identify $\tau_{\rm int}$ with the relaxation time scale of the degrees of freedom that $\phi$ couples to \cite{Berges:2005md}, i.e., the time associated with partial memory loss, which in general is considerably shorter than the time scale on which full thermal equilibrium is established. 
For $\tau_{\rm int}\flatfrac{\dot{\varphi}}{\varphi}\ll1$ one can apply the approximation \eqref{AD-approx} inside the integrals, which effectively makes the equation \eqref{FunctionalTaylor} local in time.\footnote{\label{WignerWeisskopf}
This approximation is similar to the Wigner-Weisskopf approximation \cite{Weisskopf:1930au} in atomic physics.
Here we have assumed that $\varphi$ is the only degree of freedom that is out of equilibrium to keep the derivation simple. 
If the correlation functions exhibit a time dependence from other origins than their dependence on $\varphi$ (e.g.~due to 
non-equilibrium initial conditions for the propagators $\Delta$ themselves,
time-dependent external sources, couplings to other out-of-equilibrium degrees of freedom, a thermal bath with time-dependent temperature $T$, or a time-dependent space-time metric), then an expansion of all time-dependent quantities in analogy to \eqref{AD-approx} must be performed. Our method can be applied as long as all time-dependent quantities fulfil the second inequality in \eqref{RangeOfValidity}, e.g., $\tau_{\rm int} \dot{T}/T\ll 1$ etc.
}  
This does not imply any restriction on the field excursion because the expansion can be applied locally at each moment $t=x^0$. We can therefore approximate $\delta\varphio(x_i) =  \dot{\varphi}(t)(x_i^0-t)$ to bring  the equation of motion \eqref{EOM-varphi0} into the form \eqref{EOM-slow-roll-LTE},
\begin{align}\label{Master}
  \ddot{\varphi} +
\sum_{n=1}^{+\infty} \Gamma_{\varphi}^{(n)} \dot{\varphi}^n+
\partial_{\varphi} \mathcal{V}_\varphi
  =0
\end{align}
with
\begin{eqnarray}
\partial_{\varphi} \mathcal{V}_\varphi &=& \partial_\varphi V(\varphi(t))-\left. \frac{\partial \mathbf{\Gamma}_{\rm loop}[\varphi,  \Delta[\varphi]]}{\partial \varphi(x)} \right|_{
\substack{t}
}\label{EffPot}
\end{eqnarray}
and
\begin{eqnarray}\label{Gamma_functional_taylor_exp}
  \Gamma_{\varphi}^{(n)} =  
  -\frac{1}{n!}
 \prod_{i=1}^{n}  \left[
 \int_{\mathcal{C}}  dx_i^0 (x_i^0 -t) 
 \int d^3 \x_i \right]
 \left[\frac{\delta^n}{\delta \varphi(x_1) \cdots \delta\varphi(x_{n})}\left(\frac{\partial\mathbf{\Gamma}_{\rm loop}[\varphi,  \Delta[\varphi]]}{\partial \varphi(x)}\right)\right]\Bigg|_{t}.
 \end{eqnarray}
Here it is important to note that the factor $\frac{\delta^n}{\delta \varphi(x_1) \cdots \delta\varphi(x_n)}$ is to be understood as a total functional derivative which acts both on $\varphi$ directly and also on $\Delta[\varphi]$.
The notation $|_t$ implies that all quantities must be evaluated at the reference time $t$, cf.~footnote~\ref{WignerWeisskopf}, which in particular implies $\varphi=\varphio$.

A few comments are in place.
First, it is worthwhile noting that, while the expansion used here formally enables one to include terms of all orders in $\dot{\varphi}$, it neglects higher derivatives of $\varphi$. The second derivative $\ddot{\varphi}$ will introduce a correction to the kinetic term, which in principle can be included in a straightforward way \cite{Cametti:1999ii}.
We do not consider these terms here because they correspond to a perturbative correction to a term that already exists in the classical action.
This is in contrast to the friction term, which only appears once loop corrections are included.
Higher derivatives can lead to spurious behaviour in the truncated local equations. However, if the theory at a fundamental level is well behaved, these problems can be kept under control if the derivative expansion is done carefully \cite{Glavan:2017srd}.
Second, it turns out to be crucial that we formulated the equations in the 2PI framework.
For vanishing external sources all $n$-particle irreducible ($n$PI) effective actions must give the same results for physical observables at the level of exact equations.\footnote{
The assumption that all external sources vanish is crucial for this statement to be true. If the $n$-th source is non-trivial, then the $n$PI effective action can contain more information than the $(n-1)$PI effective action. A pedagogical discussion can be found in \cite{Millington:2019nkw}, practical applications include computations in curved space-time \cite{Calzetta:1986ey} and false vacuum decay \cite{Garbrecht:2015yza}.
}  
However, at any finite order in the loop expansion, the different $n$PI effective actions correspond to different resummation and truncation schemes, and they can give different results.
As we will show explicitly in Eq.~\eqref{GammaTadpole}, the inclusion of self-energies inside the loops is crucial. This feature would not be captured in the 1PI formalism unless a resummation of the propagators is done by hand \cite{Cheung:2015iqa}.

\section{Interpretation of the leading friction term}\label{sec:LeadingTerm}
In the previous section we have derived a Markovian master equation \eqref{Master} for $\varphi$. Equations \eqref{EffPot} and \eqref{Gamma_functional_taylor_exp} provide a recipe to compute the coefficients.
The only assumption required is that the approximation \eqref{AD-approx} can be used, i.e., that $\tau_{\rm int}\flatfrac{\dot{\varphi}}{\varphi}\ll 1$.
This assumption is crucial in two ways.
First, it makes each term in the expansion \eqref{FunctionalTaylor} local in time. Second, the smallness of $\dot{\varphi}$ serves as justification to consider only a finite number of terms \eqref{Gamma_functional_taylor_exp}.
In the following we study the properties of the coefficients to gain  insight into their microphysical interpretation without resorting to a specific model.
For this purpose it is instructive to explicitly consider the truncation of \eqref{EOM-slow-roll-LTE}
that only includes the leading term $\Gamma_\varphi \equiv \Gamma_\varphi^{(1)}$, i.e.\footnote{
For small field values an effective equation of the form \eqref{EOM-hom-3} can also be obtained from
\eqref{EOM-varphi0} and \eqref{FunctionalTaylor}  
without the approximation \eqref{AD-approx} by truncating the expansion in $\delta\varphi$ at linear order \cite{Calzetta:1989vs,Morikawa:1986rp}. 
If the ground state is at $\varphi=0$, this truncation implies that propagators $\Delta$ are in good approximation independent of $\varphi$. Then \eqref{EOM-varphi0} takes the form of a Langevin equation with vanishing noise term (cf. footnote \ref{NoiseFootnote}), which can be solved by means of linear response theory \cite{GreenMelville,Kubo:1957mj}.
The linear response here implies that the 
 \eqref{EOM-slow-roll-LTE} is truncated at $n=1$ and the coefficients are independent of $\varphi$
(because the effect that $\varphi$ has on the propagators of $\phi$ and other fields is neglected), $\varphi$ is exposed to Brownian motion due to its coupling to other field modes, but does not affect those.
Corrections beyond the linear order in $\delta\varphi$ can be included by means of multi-scale perturbation theory \cite{Ai:2021gtg}.
}
\begin{equation}
\ddot{\varphi} + \Gamma_\varphi\dot{\varphi} + \partial_{\varphi} \mathcal{V} =0  .
\label{EOM-hom-3}
\end{equation}
In the present context this truncation can be justified because $\varphi$ evolves slowly, such that higher powers of $\dot{\varphi}$ can be considered subdominant.  
We compute $\Gamma_\varphi$ from the lowest order term in our expression \eqref{Gamma_functional_taylor_exp}, which was derived by using \eqref{AD-approx}, but without limiting the field excursion, and therefore is valid beyond the linear response regime. 
For the following discussion we split
\begin{equation}\label{Pi_tprime_t}
\Gamma_\varphi= \Gamma_\varphi^{[1]}+
\Gamma_\varphi^{[2]} \quad , \quad
\Gamma_\varphi^{[i]} =  
\int_{\mathcal{C}} \int d^3 \x_1 dx_1^0 (x_1^0-t) \,
\Uppi_\varphi^{[i]}(x_1,x) 
\end{equation}
with
\begin{eqnarray}
\Uppi_\varphi^{[1]}(x_1,x) &=& - \left.
\frac{\partial^2 \mathbf{\Gamma}_{\rm loop}[\varphi,  \Delta[\varphi]]}{\partial \varphi(x_1) \partial \varphi(x)}\right|_{\substack{\varphio}},\label{Pivarphi1insec2} \\   
\Uppi_\varphi^{[2]}(x_1,x) &=& - \left.\sum_{a,b} \int_y \int_z \left(\left.\frac{\partial^2 
\mathbf{\Gamma}_{\rm loop}[\varphi, \Delta]}{ \partial \Delta_{ab}(y,z) \partial\varphi(x)}\right|_{\Delta[\varphi]}\frac{\delta \Delta_{ab}[\varphi](y,z)}{\delta\varphi(x_1) }\right)\right|_{\substack{\varphio}} \,.
\label{Pi_varphi_Def}
\end{eqnarray}
This splitting makes explicit that the friction coefficient receives 
two contributions. The first term, $\Uppi_\varphi^{[1]}$ 
comes from the explicit $\varphi$-dependence of $\mathbf{\Gamma}_{\rm loop}$, which in the perturbative expansion is due to the $\varphi$-dependent vertices that appear in the action when splitting $\phi = \varphi + \eta$. 
The second term, $\Uppi_\varphi^{[2]}$, comes from the time dependence of the resummed propagators $\Delta[\varphi]$ in the loop expansion. 
Under the simplified assumptions that we made for illustrative purposes, cf.~footnote \ref{WignerWeisskopf}, the sole origin of this time dependence is the dependence of the propagators on $\varphi$ itself. Hence, $\Uppi_\varphi^{[2]}$ can be understood as the non-linear feedback from the impact that $\varphi$ has on the properties of the (quasi)particles that it interacts with on the evolution of $\varphi$ itself.

To obtain a microphsical interpretation of the dissipative term $\Gamma_\varphi$ we first consider the case where $\phi$ resides in an environment that is in thermal equilibrium, and the deviation from its ground state is small 
in the sense of linear response discussed after \eqref{EOM-hom-3}. Then the friction term $\Gamma_\varphi$ can simply be obtained from thermal field theory, cf.~e.g.~\cite{Morikawa:1986rp,Greiner:1998vd,Yokoyama:2004pf,Boyanovsky:2004dj,Anisimov:2008dz,Gautier:2012vh,Mukaida:2013xxa,Ai:2021gtg}. 
This can e.g.~be justified if the temperature  $T$ is much larger than $\varphi$, or if the bath contains sufficiently many degrees of freedom,
so that the effect of the $\varphi$-evolution on the temperature of the bath can be neglected. 
Then $\V$ and $\Gamma_\varphi$ can be related to the real and imaginary parts of retarded self-energies in thermal field theory, evaluated at multiples of the oscillation frequency, cf.~e.g.~Refs.~\cite{Boyanovsky:2004dj,Anisimov:2008dz,Cheung:2015iqa,Mukaida:2013xxa} for explicit derivations.
In this case one can apply the optical theorem at finite temperature, which relates the imaginary part of self-energies to cuts through Feynman diagrams \cite{Weldon:1983jn,Kobes:1985kc,Kobes:1986za,Landshoff:1996ta,Gelis:1997zv,Bedaque:1996af}, and $\Gamma_\varphi$ can be interpreted in terms of the creation of particles \cite{Calzetta:1989vs} and interactions with the constituents of the thermal bath medium \cite{Weldon:1983jn,Morikawa:1986rp}.
This provides an intuitive interpretation of the friction in terms of microphysical processes, as the different cuts can represent decays and scatterings amongst quasiparticles that transfer energy between the different constituents of a system \cite{Weldon:1983jn,Boyanovsky:2004dj,Drewes:2010pf,Drewes:2013iaa}.

An important consequence of the finite temperature cutting rules is that local diagrams cannot contribute to dissipation in thermal field theory. 
More generally, any contribution to $\Uppi_\varphi^{[i]}(x_1,x)$ that contains a $\delta(x_1^0-t)$ cannot contribute to $\Gamma_\varphi$ because of the convolution with $(x_1^0-t)$ in Eq.~\eqref{Pi_tprime_t}. 
Based on this one may naively expect that $\mathbf{\Gamma}_1[\varphi,  \Delta[\varphi]]$ in \eqref{2PI-action} cannot contribute to $\Gamma_\varphi$, and
dissipation solely comes from the non-local term  $\mathbf{\Gamma}_2[\varphi,  \Delta[\varphi]]$. The latter is, however, not true because in a time-dependent background the situation is more subtle. 
The reason is that 
$\mathbf{\Gamma}_1$ comprises diagrams that are not truly local:
Through the resummation effects
they include integrations over time and nested nonlocalities. 
More precisely, the statement that $\mathbf{\Gamma}_1$ does not contribute to dissipation holds under the assumption that propagators are time translation invariant.
In the simple setup considered here this assumption is violated by the non-linear feedback induced by the $\varphi$-dependence of the propagators from which the dissipation coefficients are computed, which gives rise to the additional term $\Uppi_\varphi^{[2]}$.

The $\mathbf{\Gamma}_1$ contribution to $\Uppi_\varphi^{[1]}(x_1,x_2)$ is given by 
\begin{equation}\label{Pi1piece}
    \Uppi_\varphi^{[1]}(x_1,x)\supset-\left.\frac{\partial^2 \mathbf{\Gamma}_{1}[\varphi,  \Delta[\varphi]]}{\partial \varphi(x_1) \partial \varphi(x)}\right|_{\substack{\varphio }}=\left.-\frac{i}{2}\sum_{a,b}\int_z\int_{z'}\left(\frac{\partial^2G ^{-1}_{0,ab}[\varphi](z,z')}{\partial\varphi(x_1)\partial\varphi(x)}\Delta_{ab}[\varphi](z,z')\right)\right|_{\substack{\varphio}},
\end{equation}
where partial derivatives only act on $G ^{-1}_{0,ab}[\varphi]$ as explained below Eq.~\eqref{tree_level_delta}. Using the definition of $G^{-1}_{0,ab}[\varphi](z,z')$, we find
\begin{align}\label{Pi1local}
    \frac{\partial^2i G^{-1}_{0,ab}[\varphi](z,z')}{\partial\varphi(x_1)\partial\varphi(x)}&=\frac{1}{2}\frac{\partial^2}{\partial\varphi(x_1)\partial \varphi(x)}\left(\left. \frac{\partial^2 S[\Phi]}{\partial\Phi_a(z) \partial\Phi_b(z')} \right|_{\langle \Phi \rangle}\right)\nonumber \\ 
    & \sim \delta_\mathcal{C}(z-z')\delta_\mathcal{C}(z-x_1)\delta_\mathcal{C}(z-x).
\end{align}
By putting the last two equations together, we see that the $\mathbf{\Gamma}_1$ contribution to $\Uppi_\varphi^{[1]}(x_1,x)$ ends up being proportional to $\delta_\mathcal{C}(x_1^0-t)$ which, once convolved 
with the $(x_1^0-t)$ of $\Gamma_\varphi^{[1]}$, vanishes. Thus, we can replace $\mathbf{\Gamma}_{\rm loop}$ by $\mathbf{\Gamma}_2$ in $\Uppi_\varphi^{[1]}$ and conclude that only non-local 2PI diagrams, which contain at least two separate vertices depending on the background field, can give a non-vanishing contribution to the damping rate 
$\Gamma_\varphi^{[1]}$.

However, the one-loop term $\mathbf{\Gamma}_1$ does contribute to $\Gamma_\varphi$ through $\Uppi_\varphi^{[2]}(x_1,x)$. We first note that $\Uppi_\varphi^{[2]}(x_1,x)$ in \eqref{Pi_varphi_Def} is an integral of a product of two factors. We then work out how these two factors are dealt with in practice, focusing on the $\mathbf{\Gamma}_1$ contribution. Using Eqs.~\eqref{Gamma1} and \eqref{tree_level_delta}, we find for the first factor in \eqref{Pi_varphi_Def}
\begin{equation}\label{first_factor}
\left.\frac{\partial^2 
\mathbf{\Gamma}_1[\varphi, \Delta]}{ \partial\Delta_{ab}(y,z) \partial \varphi(x)}\right|_{\Delta[\varphi]}
= \frac{1}{2}\frac{\partial}{\partial \varphi(x)}\left(\left. \frac{\partial^2 S[\Phi]}{\partial\Phi_a(y) \partial\Phi_b(z)} \right|_{\langle \Phi \rangle }\right) \sim \delta_\mathcal{C}(y-z)\delta_\mathcal{C}(y-x).
\end{equation}
With the help of Eq.~\eqref{Inverse_G}, together with the relation $\delta A = - A \delta (A^{-1}) A$ valid for small perturbations $\delta A$ around any invertible operator $A$, the second factor can be written as
\begin{align}
\frac{\delta \Delta_{ab}[\varphi](y,z)}{\delta \varphi(x_1) }
& = - \left(\Delta
\frac{\delta \Delta^{-1}}{\delta\varphi(x_1) }
\Delta\right)_{ab}[\varphi](y,z), \; \; \; \text{(in the matrix product sense)}
\nonumber \\
& = - \sum_{a'b'}\int_{y'}\int_{z'}  \Delta_{aa'}[\varphi](y,y')
\frac{\delta\left(G_{0,a'b'}^{-1}[\varphi](y',z')-\Pi_{a'b'}[\varphi](y',z') \right)}{\delta\varphi(x_1) }
\Delta_{b'b}[\varphi](z',z) \label{from_local_to_non_local}.
\end{align}
The latter is made of two terms, functional derivatives of $G_0^{-1}[\varphi]$ and of $\Pi[\varphi]$. Taking the product of them with Eq.~\eqref{first_factor} above, we see that both can lead to nonvanishing contributions to the damping rate, which originate from ``local" one-loop diagrams. 
The reason is that they include integrations over time and nested nonlocalities
through the resummation effects that we implemented in the last equation when interting the 2PI Schwinger-Dyson equation of motion.
If one interprets these terms as higher order corrections in the small field expansion within the 1PI scheme, they can be interpreted in terms of elementary processes by applying cutting rules \cite{Cheung:2015iqa}. The advantage of the approach in the present work is that it is valid beyond the small field expansion.

\section{A simple example}\label{sec:model}
\begin{figure}[h!]
\begin{center}
\includegraphics[scale=0.7]{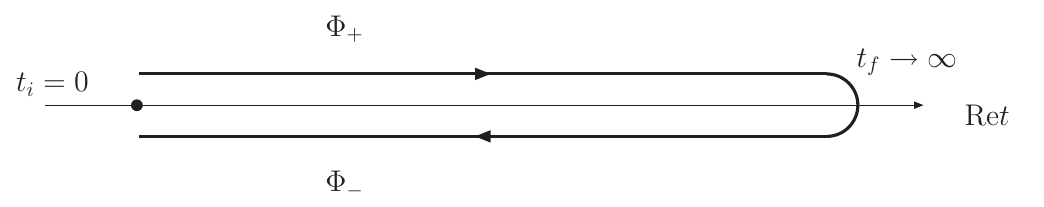}
\caption{The Closed Time-Path (CTP) of the Schwinger-Keldysh formalism}
\label{Contour-CTP}
\end{center}
\end{figure}

For illustrative purposes, and to allow for a direct comparison with the literature, we apply our method to the same $Z_2-$symmetric theory\footnote{$Z_2-$symmetric theories with only one nonvanishing expectation value are computationally very convenient because there is no mixing between the particle species and the associated resummed propagators.}
that has been studied in \cite{Cheung:2015iqa} by means of a small field expansion.
Consider the action
\begin{equation}\label{Lagrangian}
   S[\phi,\chi]= \int_x \left[
\frac{1}{2}\partial_\mu\phi\partial^\mu \phi
-\frac{1}{2}m_\phi^2\phi^2+\frac{1}{2}\partial_\mu\chi\partial^\mu \chi
-\frac{1}{2}m_{\chi}^2\chi^2 
-\frac{h}{4}\phi^2\chi^2  -\mathcal{L}_{\chi {\rm int}} 
\right],
\end{equation}
where $\phi$ is the field of nonvanishing expectation value $\varphi=\langle \phi \rangle$ with $\phi = \varphi + \eta$. We choose $\mathcal{C}$ to be the Closed Time-Path of the Schwinger-Keldysh formalism shown in Fig.~\ref{Contour-CTP}, often referred to as \emph{Closed Time-Path}  \cite{Schwinger:1960qe,Bakshi:1962dv,Bakshi:1963bn,Keldysh:1964ud}.
We want to determine  
the contributions $\Uppi_\varphi^{[1]}$ and $\Uppi_\varphi^{[2]}$ to the friction coefficient $\Gamma_\varphi$ \eqref{Pi_tprime_t}
from Eqs.~\eqref{Pivarphi1insec2} and \eqref{Pi_varphi_Def} in the model \eqref{Lagrangian}.
In order to obtain analytical results we make a number of simplifying assumptions.
We assume that $\langle\chi\rangle=0$, and that all degrees of freedom in the system except $\varphi$ reside in thermal equilibrium with temperature $T$
at all times.
This equilibrium is established by the interactions in $\mathcal{L}_{\chi {\rm int}}$ on time scales $\tau_{\rm eq}$ with
$\tau_{\rm int}\ll \tau_{\rm eq} \ll \flatfrac{\varphi}{\dot{\varphi}}$, which could e.g.~be realised through a hierarchy between $h$ and the coupling constants in $\mathcal{L}_{\chi {\rm int}}$.
We further assume that $T$ remains constant, which amounts to neglecting the effect that the dissipation of energy from $\varphi$ into particles has on the temperature. Physically this can e.g.~be realised if $\mathcal{L}_{\chi {\rm int}}$ couples $\chi$ to a sufficiently large thermal bath.
Hence, all time dependence in the system comes from the coupling to $\varphi(t)$ alone.
We emphasise that this does not restrict the validity of the analysis to the linear response regime because we make no assumption on the elongation of $\varphi$ at initial time, and we do take into account non-linear effects in $\varphi$ through the $\varphi$-dependence of the propagators $\Delta[\varphi]$.

\begin{figure}[!ht] 
\centering
\begin{tikzpicture}[scale=2.7]
\begin{scope}[shift={(-2,0)}]
\draw[color=black] (0.4,0.15) circle (10pt);
\draw[dashed,color=black] (1.1,0.15) circle (10pt);
\draw[fill=black] (0.75,0.15) circle (1pt);
\end{scope}

\begin{scope}[shift={(0,0)}]
\draw[dashed,color=black] (0.4,0.15) circle (10pt);
\draw (0.05,0.15) --  (0.75,0.15) circle (1pt);
\draw[fill=black]  (0.05,0.15) circle (1pt);
\draw[fill=black]  (0.75,0.15)circle (1pt) ;
\draw  (0.05,0.15) --(0.05,0.55);
\draw  (0.75,0.15) --(0.75,0.55);
\draw  (0.05,0.55) node[] {$\times$};
\draw  (0.75,0.55) node[] {$\times$};
\end{scope}

\begin{scope}[shift={(1.5,0)}]
\draw[color=black] (0.4,0.15) circle (10pt);
\draw[dashed] (0.4,0.15) ellipse (0.175cm and 0.35cm);
\draw[fill=black] (0.4,-0.2) circle (1pt);
\draw[fill=black] (0.4,0.5) circle (1pt);
\end{scope}
\end{tikzpicture}
\caption{2PI diagrams which contribute to $\mathbf{\Gamma}_2[\varphi, \Delta]$ at three-loop order. Solid (resp. dashed) lines represent full $\eta$ (resp. $\chi$) -propagators. Black circles and crosses respectively represent couplings $-ih$ and attachments of $\varphi$.}
\label{three-loop-diagrams}
\end{figure}
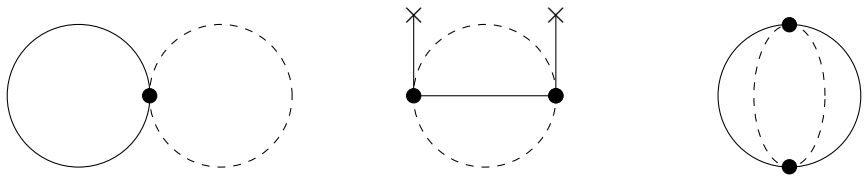

\paragraph{Computation of $\Gamma_{\varphi}^{[1]}$.}
We first start by evaluating $\Uppi_\varphi^{[1]}$ from \eqref{Pivarphi1insec2}.
As shown after Eq.~\eqref{Pi1local}, 
$\Gamma_\varphi$ does not receive a contribution from $\mathbf{\Gamma}_{1}$ through $\Uppi_\varphi^{[1]}$ because of the 
convolution with $(x_1^0-t)$ in Eq.~\eqref{Pi_tprime_t}.
We therefore focus on 
$\mathbf{\Gamma}_2[\varphi, \Delta]$ which up to three-loop level reads
\begin{eqnarray}
\mathbf{\Gamma}_2[\varphi, \Delta]&=& - \frac{i}{2^2}   \int_x \left(-i h \right) \Delta_{\eta \eta}(x,x) \Delta_{\chi \chi}(x,x)
 \nonumber \\
&& - \frac{i}{2^2} \int_{x,y}
 \left(-i h \varphi(x) \right) \left(-i h \varphi(y) \right) \Delta_{\chi \chi}^2(x,y)\Delta_{\eta \eta}(x,y) \nonumber \\
&& -\frac{i}{2^3}\int_{x,y} (-ih)^2 
\Delta_{\eta \eta}^2(x,y) \Delta_{\chi \chi}^2(x,y) 
\,. \label{Gamma_2}
\end{eqnarray}
Diagrammatically, it contains three diagrams, which we depict in Fig.~\ref{three-loop-diagrams}.

Looking at Fig.~\ref{three-loop-diagrams}, it is clear that only the setting-sun diagram with two couplings to the external background $\varphi$ can contribute because the partial functional derivative in \eqref{Pivarphi1insec2} only acts on the explicit factors $\varphi$ at the vertices, not on the implicit $\varphi$-dependence of the propagators. 
Calculating it explicitly, we obtain
\begin{equation}
    - \left.
\frac{\partial^2 \mathbf{\Gamma}_{2}[\varphi,  \Delta[\varphi]]}{\partial \varphi(x_1) \partial \varphi(x)}\right|_{\substack{\varphio}}=  -\frac{i h^2}{2} 
 \Delta_{\chi \chi}^2[\varphio](x,x_1)\Delta_{\eta \eta}[\varphio](x,x_1), 
\end{equation}
which still has to be convolved with $(x_1^0-t)$ in \eqref{Pi_tprime_t} to obtain the final contribution to $\Gamma_\varphi$. 
The loop integrals are to be evaluated with full propagators, but with $\varphi=\varphio$, i.e.,~in a (locally) static background.
To evaluate this expression it is convenient to decompose the propagator on the contour $\mathcal{C}$ as
\begin{eqnarray} \label{decomp-F-rho}
 \Delta_{\chi\chi}(x,y)= \Delta^+_{\chi\chi}(x,y)-\frac{i}{2}  \Delta^-_{\chi\chi}(x,y)\, \text{sign}_{\mathcal{C}}(x^{0}-y^{0}),
\end{eqnarray} 
where the spectral function $\Delta^-$ and statistical propagator $\Delta^+$ are defined as
\begin{eqnarray}
 \Delta_{\chi\chi}^+(x_1,x_2) &=& \frac{1}{2}
 \left(\Delta_{\chi\chi}^>(x_1,x_2) + \Delta_{\chi\chi}^<(x_1,x_2)\right)
 - \langle \chi(x_1)\rangle \langle\chi(x_2)\rangle  \,,\label{DefF} \\
 \Delta_{\chi\chi}^-(x_1,x_2) &=& i 
 \left(\Delta_{\chi\chi}^>(x_1,x_2) - \Delta_{\chi\chi}^<(x_1,x_2)\right)
 ,\label{Defrho}\\
\Delta_{\chi\chi}^>(x_1,x_2) &=& \langle\chi(x_1)\chi(x_2)\rangle  - \langle \chi(x_1)\rangle \langle\chi(x_2)\rangle \nonumber
\ , \\
 \Delta_{\chi\chi}^<(x_1,x_2) &=& \langle\chi(x_2)\chi(x_1)\rangle - \langle \chi(x_1)\rangle \langle\chi(x_2)\rangle  .\label{Wightman}
\end{eqnarray}
 On the contour in Fig.~\ref{Contour-CTP}, we see that 
 \begin{eqnarray}\label{Ruhleben}
 \int_{\mathcal{C}} dy^0 \textrm{sign}_{\mathcal{C}}^n(x^0-y^0) \ldots
 =
 \begin{cases}
  0 \ {\rm for } \ n \ {\rm even}\\
  2 \int_{t_i}^{x^0 }  dy^0 \ldots \ {\rm for } \ n \ {\rm odd}
 \end{cases},
 \end{eqnarray}
because the second part of the contour runs backwards (i.e., anti-parallel to the real axis), and  times on the second part are always later than those on the first.
We further notice that the contour propagators with time arguments on different branches of the contour can be identified with the Wightman functions $\Delta^\gtrless$ in \eqref{Wightman}. This property also allows us to define self-energies $\Pi^\gtrless$ in the same way. 
 Since we are dealing with a homogeneous and isotropic system, all quantities can only depend on the difference between the spatial coordinates, and we can perform a Fourier transform in this difference.
 Using \eqref{decomp-F-rho}, \eqref{DefF}, \eqref{Defrho} and \eqref{Ruhleben} we can write 
\begin{eqnarray}
\Gamma_{\varphi}^{[1]}&=&- i h^2\int_{t_i}^t dx_1^0\,(x_1^0-t)\int \frac{d^3\p}{(2\pi)^3}\frac{d^3\q}{(2\pi)^3}\,\Bigg[
\Delta_{\chi\chi}^>[\varphi](t,x_1^0;\p)
\Delta_{\chi\chi}^>[\varphi](t,x_1^0;\q)
\Delta_{\eta\eta}^>[\varphi](t,x_1^0;-\q-\p)
\nonumber\\
&&- \Delta_{\chi\chi}^<[\varphi](t,x_1^0;\p)
\Delta_{\chi\chi}^<[\varphi](t,x_1^0;\q)
\Delta_{\eta\eta}^<[\varphi](t,x_1^0;-\q-\p)
\Bigg]_{\varphio}\\
&=&
- i h^2\int_{0}^{t-t_i} dz z\int \frac{d^3\p}{(2\pi)^3}\frac{d^3\q}{(2\pi)^3}\,\Bigg[
\Delta_{\chi\chi}^>[\varphi](t,t-z;\p)
\Delta_{\chi\chi}^>[\varphi](t,t-z;\q)
\Delta_{\eta\eta}^>[\varphi](t,t-z;-\q-\p)
\nonumber\\
&&- \Delta_{\chi\chi}^<[\varphi](t,t-z;\p)
\Delta_{\chi\chi}^<[\varphi](t,t-z;\q)
\Delta_{\eta\eta}^<[\varphi](t,t-z;-\q-\p)
\Bigg]_{\varphio}
\end{eqnarray}
where we have defined $z=t-x^0_1$ in the last line. Explicit expressions for $\Delta^+$ and $\Delta^-$ in an adiabatically changing background have e.g.~been obtained in Ref.~\cite{Drewes:2012qw}. 
However, since we evaluate the integrand in \eqref{Pi2approx} at $\varphi=\varphio$, only the static limit is needed,\footnote{As pointed out before,
we can consider $\varphio$ as static under the integral because the integrand is suppressed for time separations  $|z| \gg \tau_{\rm int}$, practically acting as a window function.
However, this is of course a moving window, as $\varphio$ is fixed to the value of $\varphi$ at the reference time $t$, which itself is dynamical.
} 
which has e.g. been computed in Ref.~\cite{Anisimov:2008dz}.
In the static limit $\Delta^+$ and $\Delta^-$ are only functions of the relative time coordinate $z$, which we express by a slight abuse of notation,\footnote{
We remind the reader that we have assumed a constant temperature $T$ for simplicity and illustrative purposes. An adiabatic change in $T$ can be treated systematically with the method presented here, cf.~footnote \ref{WignerWeisskopf}.
}
\begin{eqnarray}\label{珍珠奶茶}
\Gamma_{\varphi}^{[1]}&=&
- i h^2\int_{0}^{t-t_i} dz z\int \frac{d^3\p}{(2\pi)^3}\frac{d^3\q}{(2\pi)^3}\,\Bigg[
\Delta_{\chi\chi}^>[\varphio](z;\p)
\Delta_{\chi\chi}^>[\varphio](z;\q)
\Delta_{\eta\eta}^>[\varphio](z;-\q-\p)
\nonumber\\
&&- \Delta_{\chi\chi}^<[\varphio](z;\p)
\Delta_{\chi\chi}^<[\varphio](z;\q)
\Delta_{\eta\eta}^<[\varphio](z;-\q-\p)
\Bigg]
\end{eqnarray}
For $|t-t_i|\gg \tau_{\rm int}$ we can take the limit
$t_i\to-\infty$. 
This limit is not strictly necessary to obtain a Markovian equation, which was already achieved by the replacement $\varphi\to\varphio$ in \eqref{珍珠奶茶}, but it simplifies the computation of the dissipation coefficient and implies no additional approximation (since the validity of our method is in any case restricted to times longer than $\tau_{\rm int}$).
From the symmetry of the Wightman functions $\Delta^\gtrless$ it follows that the integrant is symmetric in $z$, so that we can write 
\begin{eqnarray}
\Gamma_{\varphi}^{[1]}&=&
- \frac{i}{2} h^2\int_{-\infty}^{\infty} dz z\int \frac{d^3\p}{(2\pi)^3}\frac{d^3\q}{(2\pi)^3}\,\Bigg[
\Delta_{\chi\chi}^>[\varphio](z;\p)
\Delta_{\chi\chi}^>[\varphio](z;\q)
\Delta_{\eta\eta}^>[\varphio](z;-\q-\p)
\nonumber\\
&&- \Delta_{\chi\chi}^<[\varphio](z;\p)
\Delta_{\chi\chi}^<[\varphio](z;\q)
\Delta_{\eta\eta}^<[\varphio](z;-\q-\p)
\Bigg]\\
&=&
- \frac{i}{2} h^2\lim_{\omega\to 0} \frac{\partial}{i\,\partial\omega}\int_{-\infty}^{\infty} dz 
\int \frac{d^3\p}{(2\pi)^3}\frac{d^3\q}{(2\pi)^3}\,e^{i\omega z}\Bigg[
\Delta_{\chi\chi}^>[\varphio](z;\p)
\Delta_{\chi\chi}^>[\varphio](z;\q)
\Delta_{\eta\eta}^>[\varphio](z;-\q-\p)
\nonumber\\
&&- \Delta_{\chi\chi}^<[\varphio](z;\p)
\Delta_{\chi\chi}^<[\varphio](z;\q)
\Delta_{\eta\eta}^<[\varphio](z;-\q-\p)
\Bigg].\label{GainLoss1}
\end{eqnarray}
By performing a Fourier transformation in $z$
and using the relation ${\rm Im}\Pi^R(p)=\frac{1}{2}(\Pi^>(p) - \Pi^<(p))$
this integral can be related to the contribution to the retarded $\phi$-self-energy from the ``setting sun diagram"  obtained by applying \eqref{PiDef} to the third diagram in Fig.~\ref{three-loop-diagrams},
\begin{eqnarray}
\Gamma_{\varphi}^{[1]}
&=&
- \frac{i}{2} \lim_{\omega\to 0} \frac{\partial}{i\,\partial\omega}
\left(
\Pi^>_{\rm sun}[\varphio](\omega;\textbf{0}) - \Pi^<_{\rm sun}[\varphio](\omega;\textbf{0})
\right)\nonumber\\
&=&
- \lim_{\omega\to 0} \frac{\partial}{\partial\omega}
{\rm Im}\Pi^R_{\rm sun}[\varphio](\omega;\textbf{0}) = 
- \lim_{\omega\to 0} \frac{{\rm Im}
\Pi^R_{\rm sun}[\varphio](\omega;\textbf{0})
}{\omega},\label{SonsOfCain}
\end{eqnarray}
in agreement with the general observation that dissipative behaviour is associated with the imaginary parts (more precisely: discontinuities) of retarded self-energies \cite{Weldon:1983jn}.
This diagram has been estimated before in Ref.~\cite{Cheung:2015iqa} with the approximation \eqref{AD-approx},  the resulting contribution to $\Gamma_\varphi$ reads 
\begin{eqnarray}\label{Gammasun_large_Meta}
\Gamma_\varphi^{\rm [1]}\simeq\frac{ h^2 \,T^2}{  (4 \pi)^3 {M}_{\eta}} \log\left(\frac{{M}_{\eta}}{{M}_{\chi}}\right)  \;  {\rm for} \; T \gg {M}_\eta \gg {M}_{\chi},\label{Gammasun_small_Meta}
\end{eqnarray} 
where $M_a^2=m_a^2 + \delta_{a\chi}\frac{h}{2}\varphio ^2 + \frac{h}{24}T^2$.
This term can be interpreted as dissipation from scatterings with $\chi$-quanta in the thermal bath
by making connection to thermal field theory \cite{Parwani:1991gq,Drewes:2013iaa}.
As expected, the friction coefficient grows with $T^2$ due to the larger number of scattering partners at higher temperature.

\paragraph{Computation of $\Gamma_{\varphi}^{[2]}$.}
We now move on to compute the contribution to $\Gamma_\varphi$ from $\Uppi_\varphi^{[2]}$, which includes contributions from both, $\mathbf{\Gamma}_{1}$ and $\mathbf{\Gamma}_{2}$. Starting from \eqref{Pi_varphi_Def}, we first evaluate the functional derivative of the propagator $\Delta$ with respect to $\varphi$ using the same technique as for Eq.\eqref{from_local_to_non_local}
\begin{equation}\label{PropExpansion}
\frac{\delta \Delta[\varphi]}{\delta \varphi(x_1)} =
- \Delta[\varphi] \frac{\delta \Delta^{-1}[\varphi]}{\delta \varphi(x_1)}\Delta[\varphi] =
- \Delta[\varphi] \frac{\delta\left(G_0^{-1}- \Pi \right)}{\delta \varphi(x_1)}\Delta[\varphi]
\simeq 
- \Delta[\varphi] \frac{\delta\left(G_0^{-1} \right)}{\delta \varphi(x_1)}\Delta[\varphi]\,,
\end{equation}
where in the last step we kept only the leading  tree-level contribution, which dominates the one of the self-energy because $\varphi$ appears with additional powers of $h$ and loop factors in the self-energy ($\left|\flatfrac{\delta\left(G_0^{-1} \right)}{\delta \varphi(x_1)}\right| > \left|\flatfrac{\delta \Pi}{\delta \varphi(x_1)}\right| $). Using this result, we calculate at leading order in the coupling constant $h$
\begin{align}\label{Torfmoorholm}
\lefteqn{\int d^3 \x_1\Pi_\varphi^{[2]}(x_1,x)=- \int d^3 \x_1  \left[
\sum_{a,b}\int_{y,z} \left.\frac{\partial^2 
\mathbf{\Gamma}_{\rm loop}[\varphi, \Delta]}{\partial \varphi(x) \partial \Delta_{ab}(y,z)}\right|_{\Delta[\varphi]}\frac{\delta \Delta_{ab}(y,z)}{\delta \varphi(x_1) }\right]_{\varphio} \,}\nonumber\\
\simeq  \int d^3 \x_1 & \int_{y,z}\Bigg[ \left.\frac{\partial^2 
\mathbf{\Gamma}_{\rm loop}[\varphi, \Delta]}{\partial \varphi(x) \partial \Delta_{\chi\chi}(y,z)}\right|_{\Delta[\varphi]}\int_{u,v}\Delta_{\chi\chi}[\varphi](y,u) \frac{\delta\left(G_{0,\chi\chi}^{-1}[\varphi](u,v) \right)}{\delta \varphi(x_1)}\Delta_{\chi\chi}[\varphi](v,z)\Bigg]_{\varphio}\nonumber\\
&= ih\varphio \int d^3 \x_1\, \int_{y,z}\left[  \left.\frac{\partial^2 
\mathbf{\Gamma}_{\rm loop}[\varphi, \Delta]}{\partial \varphi(x) \partial \Delta_{\chi\chi}(y,z)}\right|_{\Delta[\varphi]} \Delta_{\chi\chi}[\varphi](y,x_1)\Delta_{\chi\chi}[\varphi](x_1,z)\right]_{\varphio}.
\end{align}
Here we have used the explicit expression 
\begin{eqnarray} \label{G0_eta}
i G_{0, \eta \eta}^{-1}(x,y) &=& \left.\frac{\delta^2 S[\phi,\chi]}{\delta\phi(x) \delta\phi(y)}\right|_{(\phi, \chi)=(\varphi,0)} 
= -\left(\Box_x +m_\phi^2 \right) \delta_\mathcal{C}(x-y), \\
i G_{0, \chi \chi}^{-1}[\varphi](x,y) &=& \left.\frac{\delta^2 S[\phi,\chi]}{\delta\chi(x) \delta\chi(y)}\right|_{(\phi, \chi)=(\varphi,0)} 
= -\left(\Box_x +M^{\rm tree}_\chi(x)^2 \right)  \delta_\mathcal{C}(x-y) \, ,
\end{eqnarray}
where $M^{\rm tree}_\chi(x)= \sqrt{m_{\chi}^2+\frac{h}{2} \varphi(x)^2}$ and only $i G_{0, \chi \chi}^{-1}[\varphi](x,y) $ depends on $\varphi$. $\delta_\mathcal{C}$ is the four-dimensional delta function on the closed time-contour.

Now we separately study the contributions from $\mathbf{\Gamma}_{1}$ and $\mathbf{\Gamma}_{2}$ to $\Uppi_\varphi^{[2]}$, and we shall argue that the one of  $\mathbf{\Gamma}_{2}$ is suppressed compared to that of $\mathbf{\Gamma}_{1}$. Indeed, we compute
\begin{equation}\label{Gamma1Deriv}
   \left[\left. \frac{\partial^2 
\mathbf{\Gamma}_{1}[\varphi, \Delta]}{\partial \varphi(x) \partial \Delta_{ab}(y,z)}\right|_{\Delta[\varphi]}\right]_{\varphio} = -\frac{h}{2} \varphio \delta_{ab}\delta_{b\chi} \delta_{\mathcal{C}}(y-z)\delta_{\mathcal{C}}(x-y),
\end{equation}
and
\begin{eqnarray}
    \left[\left. \frac{\partial^2 
\mathbf{\Gamma}_{2}[\varphi, \Delta]}{\partial \varphi(x) \partial \Delta_{ab}(y,z)}\right|_{\Delta[\varphi]}\right]_{\varphio} &=& \frac{i h^2\varphio}{4}\Big[\delta_\mathcal{C}(x-y) + \delta_{\mathcal{C}}(x-z)\Big]\times\\
&&\times \ \Big[2 \Delta_{\chi\chi} (y,z) \Delta_{\eta\eta}(y,z) \delta_{a\chi}\delta_{ab} + \Delta^2_{\chi\chi} (y,z)\delta_{a\eta}\delta_{ab})\Big]\nonumber,
\end{eqnarray}
from where it can be seen that the piece from $\mathbf{\Gamma}_{1}$ is dominant from both, the loop and coupling constant expansions viewpoints. Therefore, we approximate
\begin{align}\label{Pi2approx}
\int d^3 \x_1\Pi_\varphi^{[2]}(x_1,x)&\simeq ih\varphio\int d^3 \x_1  \left[
\int_{y,z} \left.\frac{\partial^2 
\mathbf{\Gamma}_{1}[\varphi, \Delta]}{\partial \varphi(x) \partial \Delta_{\chi\chi}(y,z)}\right|_{\Delta[\varphi]}\Delta_{\chi\chi}[\varphi](y,x_1)\Delta_{\chi\chi}[\varphi](x_1,z)\right]_{\varphio}.
\end{align}
Again performing a Fourier transform in the relative space coordinates, we obtain
\begin{align}\label{Lienland}
-  \frac{ih^2\varphio^2}{2}\int d^3 \x_1\,  &\Bigg[\int \frac{d^3\p}{(2\pi)^3}\frac{d^3\q}{(2\pi)^3} \Delta _{\chi\chi}[\varphi](t,x_1^0;\p)\Delta_{\chi\chi}[\varphi](x_1^0,t;\q) e^{-i\p(\x-\x_1)}e^{i\q(\x-\x_1)}\Bigg]_{\varphio}\nonumber\\
&=-\frac{ih^2\varphio^2}{2}\int \frac{d^3\p}{(2\pi)^3}\,\Bigg[\Delta_{\chi\chi}[\varphi](t,x_1^0;\p)^2\Bigg]_{\varphio} .
\end{align}
Using \eqref{decomp-F-rho} and \eqref{Ruhleben}
 we obtain
\begin{align}
\Gamma_{\varphi}^{[2]}&=- 2 \times \frac{h^2\varphio^2}{2}\int_{t_i}^t dx_1^0\,(x_1^0-t)\int \frac{d^3\p}{(2\pi)^3}\,\Bigg[\Delta_{\chi\chi}^+[\varphi](t,x_1^0;\p)\Delta_{\chi\chi}^-[\varphi](t,x_1^0;\p)\Bigg]_{\varphio}\nonumber\\
&=h^2\varphio^2\int_{0}^{t-t_i} dz\,z\int \frac{d^3\p}{(2\pi)^3}\Bigg[\Delta_{\chi\chi}^+[\varphi](t,t-z;\p)\Delta_{\chi\chi}^-[\varphi](t,t-z;\p)\Bigg]_{\varphio},\label{Gamma2intermsofpropagators}
\end{align}
Following the same steps as in the computation of $\Gamma_{\varphi}^{[2]}$ we find
\begin{align}
\Gamma_{\varphi}^{[2]}&=h^2\varphio^2\int_{0}^\infty dz\,z\int \frac{d^3\p}{(2\pi)^3}\Delta_{\chi\chi}^+[\varphio](z;\p)\Delta_{\chi\chi}^-[\varphio](z;\p)\nonumber \\
&=\frac{h^2\varphio^2}{2}\int_{-\infty}^\infty dz\,z\int\frac{d^3\p}{(2\pi)^3}\Delta_{\chi\chi}^+[\varphio](z;\p)\Delta_{\chi\chi}^-[\varphio](z;\p)\nonumber\\
&=\frac{h^2\varphio^2}{2}\lim_{\omega\to 0} \frac{\partial}{i\,\partial\omega}\int_{-\infty}^\infty dz e^{i\omega z}\int \frac{d^3\p}{(2\pi)^3}\Delta_{\chi\chi}^+[\varphio](z;\p)\Delta_{\chi\chi}^-[\varphio](z;\p),\label{LucaAgr}
\end{align}
where we have again sent $t_i\to-\infty$ by assuming $|t-t_i|\gg \tau_{\rm int}$.
By means of \eqref{DefF} and \eqref{Defrho} the integrand can again be expressed in terms of Wightman functions and,
in analogy to \eqref{SonsOfCain}, one finds 
\begin{eqnarray}
\Gamma_{\varphi}^{[2]}= 
- \lim_{\omega\to 0} \frac{{\rm Im}
\Pi^R_{\rm crab}[\varphio](\omega;\textbf{0})
}{\omega},
\end{eqnarray}
with $\Pi^R_{\rm crab}$ the contribution to the retarded self-energy obtained by applying \eqref{PiDef} to the second diagram in Fig.~\ref{three-loop-diagrams}.
In the (locally) static limit we can Fourier transform \eqref{LucaAgr} with respect to $z$,
	\begin{align}
	\Gamma_{\varphi}^{[2]} &=\frac{h^2\varphio^2}{2}\lim_{\omega\to 0}\,\frac{\partial}{i\partial\omega}\int_{-\infty}^\infty dz\, e^{i\omega z}\,\int \frac{d^3\p}{(2\pi)^3} \times \nonumber\\
	    & \qquad \qquad \qquad \qquad \times \int\frac{d\omega'}{2\pi}\int \frac{d\omega''}{2\pi}e^{-i\omega' z}\,e^{-i\omega'' z}\Delta_{\chi\chi}^{+}[\varphio](\omega';\p)\Delta_{\chi\chi}^{-}[\varphio](\omega'';\p)\nonumber \\
&=\frac{h^2\varphio^2}{2}\lim_{\omega\to 0}\,\frac{\partial}{i\partial\omega} \,\int \frac{d^3\p}{(2\pi)^3}\int \frac{d\omega'}{2\pi}\Delta_{\chi\chi}^{+}[\varphio](\omega';\p)\,\Delta_{\chi\chi}^{-}[\varphio](\omega-\omega';\p)\nonumber\\
&=\frac{h^2\varphio^2}{2}\lim_{\omega\to 0}\,\frac{\partial}{i\partial\omega} \,\int \frac{d^4 p}{(2\pi)^4}\Delta_{\chi\chi}^{+}[\varphio](p_0;\p)\,\Delta_{\chi\chi}^{-}[\varphio](\omega-p_0;\p)\nonumber\\
&=-\frac{h^2\varphio^2}{2}\lim_{\omega\to 0}\,\frac{\partial}{\partial\omega} \,\int \frac{d^4 p}{(2\pi)^4}\left(\frac{1}{2}+f_B(p_0)\right)\Delta_{\chi\chi}^{-}[\varphio](p_0;\p)\,\Delta_{\chi\chi}^{-}[\varphio](\omega-p_0;\p)\nonumber\\
&=-\frac{h^2\varphio^2}{4}\lim_{\omega\to 0}\,\frac{\partial}{\partial\omega} \,\int \frac{d^4 p}{(2\pi)^4}\Big(1+f_B(p_0)
+ f_B(\omega-p_0)\Big)\Delta_{\chi\chi}^{-}[\varphio](p_0;\p)\,\Delta_{\chi\chi}^{-}[\varphio](\omega-p_0;\p)
\,\nonumber\\
&= -\frac{h^2\varphio^2}{4} \,\int \frac{d^3 \p}{(2\pi)^3}\lim_{\omega\to 0}\,\frac{\partial}{\partial\omega}\underbrace{\int \frac{d p_0}{2\pi}\Big(f_B(p_0)
- f_B(p_0-\omega)\Big)\Delta_{\chi\chi}^{-}[\varphio](p_0;\p)\,\Delta_{\chi\chi}^{-}[\varphio](\omega-p_0;\p)}_{\equiv \; I(\p)}.\label{Gamma-varphi-1loop-con}
\end{align}
In the last steps we have used that the propagators in a static background fulfill the Kubo-Martin-Schwinger relation
\begin{eqnarray}
\Delta^+(p) &=& -i\left(\frac{1}{2} + f_B(p_0)\right)\Delta^-(p)
\end{eqnarray}
where $f_B$ is the Bose-Einstein distribution which satisfies $1+ f_B(\omega)=-f_B(-\omega)$. 
To evaluate the integral we need the 
explicit expression for the spectral function in a static background \cite{Anisimov:2008dz}, 
\begin{eqnarray}\label{DeltaMinusT}
\Delta^-(p) &=& \frac{- 2 i {\rm Im}\Pi^R(p) + 2 i p_0 \epsilon}{\left(
p^2 - (M_\chi^{\rm tree})^2 - {\rm Re}\Pi^R(p)
\right)^2
+
\left(
{\rm Im}\Pi^R(p) + p_0\epsilon
\right)^2
}.
\end{eqnarray}
The retarded self-energy is defined as $\Pi^R(x_1,x_2)=\theta(t_1 - t_2)\Pi^-(x_1,x_2)$. In absence of \emph{luons} \cite{Drewes:2013bfa} and other collective excitations or bound states, the spectral function \eqref{DeltaMinusT} has four poles that we denote by $\pm\hat{\Omega}_\chi$ and $\pm\hat{\Omega}_\chi^*$. Here we have suppressed the dependence of $\hat{\Omega}_\chi$ on the spatial momentum $\textbf{p}$ for notational simplicity. 
We now shall compute the $p_0$ integral 
under the assumption that it is dominated by the pole regions by means of the Cauchy residue theorem.\footnote{
The function $\Delta^-(p)$ inherits the branch cut that $\Pi^R(p)$ has across the real $p_0$ axis, and which defines ${\rm Im}\Pi^R(p)=(\Pi^R(p_0+i\epsilon,\textbf{p}) - \Pi^R(p_0-i\epsilon,\textbf{p}) )/(2i)$.
This in principle makes the application of Cauchy's theorem tricky. 
Under the assumption that the integral is dominated by the pole regions the problem can be avoided by approximating the integrand by a Breit-Wigner function with peaks of width $\Gamma_\chi\equiv -2{\rm Im}\hat{\Omega}_\chi$ at locations $p_0=\pm\Omega_\chi \equiv\pm {\rm Re}\hat{\Omega}_\chi$. Since the Breit-Wigner function exhibits no branchcuts, it is then straightforward to apply Cauchy's theorem. 
In practice $\Omega_\chi$ can be estimated by solving  $0 = p^2 - (M_\chi^{\rm tree})^2 - {\rm Re}\Pi^R(p)$ for $p_0$, and one can approximate  $\Gamma_\chi\simeq -{\rm Im}\Pi^R(p)/p_0|_{p_0=\Omega_\chi}$; in the presence of multiple poles $\Omega_\chi^{[i]}$ from bound states of collective excitations the procedure can be applied to each pole, but their contributions to the spectral function as well as their widths have to be weighted with residues $\mathcal{Z}=1/\left(
1 -
\frac{1}{2p_0}\frac{\partial{\rm Re}\Pi^R(p)}{\partial p_0}
\right)|_{p_0=\Omega_\chi^{[i]}}$.
This procedure is only valid as long as the loop integrals are dominated by the pole regions, which physically means that the dominant contribution to all quantities of interest comes from elementary processes between on-shell quasiparticles at leading order in the loop expansion. 
There are well-known examples when this is not fulfilled, e.g., when off-shell transport \cite{Drewes:2010pf,BasteroGil:2012cm,Drewes:2013iaa} or multiple scatterings \cite{Anisimov:2010gy,Drewes:2010pf,Garbrecht:2013urw,Drewes:2013iaa,Drewes:2015eoa} play a role.}
In the following calculation it is convenient to introduce  \begin{eqnarray}
F(p_0,\omega)&=& 4 \Big(f_B(p_0)-f_B(p_0-\omega)\Big)
\textrm{Im}\Pi^R_{\chi\chi}[\varphio](p_0) \,
\textrm{Im}\Pi^R_{\chi\chi}[\varphio](\omega-p_0) \\
&=& - 4 \omega\, f_B'(p_0) \, 
\left(\textrm{Im}\Pi^R_{\chi\chi}[\varphio](p_0)\right)^2 +\mathcal{O}(\omega^2) \,
\end{eqnarray}
and 
\begin{eqnarray}
G(p_0) &=& \lim_{\omega \to 0} 
\frac{F(p_0,\omega)}{\omega}
= 
\lim_{\omega \to 0} \frac{F(-p_0,\omega)}{\omega} = \lim_{\omega \to 0} 
\frac{F(p_0+\omega,\omega)}{\omega} \nonumber \\
&=& - 4
f'_B(p_0) \left(\textrm{Im}\Pi^R_{\chi\chi}[\varphio](p_0)\right)^2
= \frac{2 \left(\textrm{Im}\Pi^R_{\chi\chi}[\varphio](p_0)\right)^2}{ T (\cosh(p_0/T)-1)} \,.  \label{Gdef}
\end{eqnarray}
Note that $G(p_0) = G(- p_0)$.
We now compute $I(\p)$ at linear order in small $\omega$ that is needed to evaluate Eq.\eqref{Gamma-varphi-1loop-con}, which will thus set all  contributions of order $\mathcal{O}(\omega^2)$ to zero eventually.
We find
\begin{align}
& -I(\p) = \int \frac{d p_0}{2 \pi} 
\frac{F(p_0, \omega)}
{(p_0^2-\hat{\Omega}_\chi^2)(p_0^2-\hat{\Omega}_\chi^{*2}){((\omega-p_0)^2-\hat{\Omega}_\chi^2)((\omega-p_0)^2-\hat{\Omega}_\chi^{*2})}} \nonumber \\
= & \frac{F(\hat{\Omega}^*_\chi,\omega)}{
4 \omega \hat{\Omega}^*_\chi\Omega_\chi 
\Gamma_\chi (2 \hat{\Omega}^*_\chi-\omega)
(2 \Omega_\chi - \omega) 
(\omega-i \Gamma_\chi  )} 
+ \frac{F(-\hat{\Omega}_\chi,\omega)}{
4 \omega \hat{\Omega}_\chi \Omega_\chi 
\Gamma_\chi (2 \hat{\Omega}_\chi + \omega)
(2 \Omega_\chi + \omega) 
(\omega -i \Gamma_\chi)} \nonumber  \\
& + \frac{F(\hat{\Omega}^*_\chi+\omega,\omega)}{
4 \omega \hat{\Omega}^*_\chi\Omega_\chi 
\Gamma_\chi (2 \hat{\Omega}^*_\chi+\omega)
(2 \Omega_\chi + \omega) 
(i \Gamma_\chi + \omega )}
+ \frac{F(-\hat{\Omega}_\chi+\omega,\omega)}{
4  \omega \hat{\Omega}_\chi \Omega_\chi 
\Gamma_\chi (2 \hat{\Omega}_\chi- \omega)
(2 \Omega_\chi - \omega) 
(i \Gamma_\chi + \omega )} 
\\
=&
\frac{(G(\hat{\Omega}^*_\chi)+\mathcal{O}(\omega))}{4 \hat{\Omega}^*_\chi \Omega_\chi \Gamma_\chi}\left(
\frac{1}{(2 \hat{\Omega}^*_\chi+\omega)
(2 \Omega_\chi + \omega) 
(i \Gamma_\chi + \omega )}- (\omega \to - \omega)\right)  \nonumber \\
& +
\frac{(G(\hat{\Omega}_\chi)+\mathcal{O}(\omega))}{4 \hat{\Omega}_\chi \Omega_\chi \Gamma_\chi}\left(
\frac{1}{(2 \hat{\Omega}_\chi+\omega)
(2 \Omega_\chi + \omega) 
(- i \Gamma_\chi + \omega )}- (\omega \to - \omega)\right) \\
=&
\frac{(G(\hat{\Omega}^*_\chi)+\mathcal{O}(\omega))}{2 \hat{\Omega}^*_\chi \Omega_\chi \Gamma_\chi}\left(
\frac{\left(2 i \Omega_\chi \Gamma_\chi + 
2 i \hat{\Omega}^*_\chi \Gamma_\chi + 4 \hat{\Omega}^*_\chi
\Omega_\chi\right) \omega + \mathcal{O}(\omega^2)
}{(4 \hat{\Omega}_\chi^{*2}-\omega^2)
(4 \Omega_\chi^2 - \omega^2) 
(\Gamma_\chi^2 + \omega^2 )}\right) + \Big(\hat{\Omega}^*_\chi \to \hat{\Omega}_\chi,  i\Gamma_\chi \to - i\Gamma_\chi\Big) \\
=& 
\frac{G(\hat{\Omega}^*_\chi) \left( i \Omega_\chi \Gamma_\chi + 
 i \hat{\Omega}^*_\chi \Gamma_\chi + 2 \hat{\Omega}^*_\chi
\Omega_\chi\right) 
}{16 \hat{\Omega}_\chi^{*3}  \Omega_\chi^3 \Gamma_\chi^3}\,
\omega + \Big(\hat{\Omega}^*_\chi \to \hat{\Omega}_\chi,  i\Gamma_\chi \to - i\Gamma_\chi\Big)  \omega+ \mathcal{O}(\omega^2)
\,.
\end{align}
In the last step we have expanded the expression for small $\omega$.
This yields
\begin{align}
\lefteqn{\lim_{\omega \to 0} \frac{\partial}{\partial \omega}I(\p)=-\frac{G(\hat{\Omega}^*_\chi) \left( i \Omega_\chi \Gamma_\chi + 
 i \hat{\Omega}^*_\chi \Gamma_\chi + 2 \hat{\Omega}^*_\chi
\Omega_\chi\right) 
}{16 \hat{\Omega}_\chi^{*3}  \Omega_\chi^3 \Gamma_\chi^3}
 + \Big(\hat{\Omega}^*_\chi \to \hat{\Omega}_\chi,  i\Gamma_\chi \to - i\Gamma_\chi\Big)} \\
&=-\frac{ \left(\textrm{Im}\Pi^{\textrm{R}}_{\chi\chi}[\varphio](\hat{\Omega}^*_\chi)\right)^2}{ T (\cosh(\hat{\Omega}^*_\chi/T)-1)}
\frac{ \left( i \Omega_\chi \Gamma_\chi + 
 i \hat{\Omega}^*_\chi \Gamma_\chi + 2 \hat{\Omega}^*_\chi
\Omega_\chi\right) 
}{8 \hat{\Omega}_\chi^{*3} \Omega_\chi^3 \Gamma_\chi^3} 
+\Big(\hat{\Omega}^*_\chi \to \hat{\Omega}_\chi,  i\Gamma_\chi \to - i\Gamma_\chi\Big) \\
&\simeq  
-\frac{ \left(\Omega_\chi \Gamma_\chi\right)^2}{ T (\cosh(\Omega_\chi/T)-1)}
\frac{ \left( 2 i \Omega_\chi \Gamma_\chi + 
  2 \Omega_\chi
\Omega_\chi\right) 
}{8 \Omega_\chi^3 \Omega_\chi^3 \Gamma_\chi^3} +\Big( i\Gamma_\chi \to - i\Gamma_\chi\Big)
 \\
&\simeq -\frac{1}{2 T \Omega_\chi^2 \Gamma_\chi (\cosh(\Omega_\chi/T)-1)} \,.
\end{align}
In the last two steps we used the Breit-Wigner approximation $\textrm{Im}\Pi_{\chi\chi}^{\textrm{R}}(\hat{\Omega}_\chi) \simeq - \Omega_\chi \Gamma_\chi $ and replaced
$\hat{\Omega}^*=\Omega$.
Using the above result in Eq.~\eqref{Gamma-varphi-1loop-con} we obtain\footnote{
The parametric dependence in this result is consistent with what has previously been found in Refs.~\cite{Berera:1998gx,Yokoyama:1998ju,Cheung:2015iqa}, but the prefactors are somewhat different.
}  
\begin{align}\label{GammaTadpole}
\Gamma_\varphi^{[2]}\simeq
\frac{h^2\varphi^2(t)}{8T }\int \frac{d^3 \p}{(2 \pi)^3}  \frac{1}{ \Omega_\chi^2 \Gamma_\chi (\cosh(\Omega_\chi/T)-1)} 
=
\frac{h^2\varphi^2(t)}{4T }\int \frac{d^3 \p}{(2 \pi)^3}  \frac{f_B(\Omega_\chi) ( 1 + f_B(\Omega_\chi) )}{ \Omega_\chi^2 \Gamma_\chi} 
\, ,
\end{align}
where we have replaced  $\varphio$ by the value $\varphi(t)$ that fixes it locally.
A further evaluation of the integral would require knowledge of the functional dependence of $\Omega_\chi$ and $\Gamma_\chi$ on $\textbf{p}$, which strongly depends on the details of the interactions $\mathcal{L}_{\chi {\rm int}}$.
We can, however, already understand many properties of the contributions $\Gamma_\varphi^{[1]}$ and $\Gamma_\varphi^{[2]}$ from Eqs.~\eqref{Gammasun_small_Meta} and \eqref{GammaTadpole}.

\paragraph{Parametric dependence and interpretation.}
As already discussed after Eq.~\eqref{Pi_varphi_Def}, $\Gamma_\varphi^{[1]}$ originates from non-local diagrams that also appear in the linear regime, i.e., when neglecting the feedback of $\varphi$ on the propagators. By using finite temperature cutting rules, these diagrams can be interpreted in terms of elementary processes, in particular scatterings with quasiparticles in the plasma. The rate at which these scatterings occur grows with temperature because the density of scattering partners increases with $T$, and also because the effect of induced transitions leads to an enhancement of the rate when the final states are highly occupied. In Eq.~\eqref{GainLoss1} this can be seen by noting that the  Kubo-Martin-Schwinger relation in thermal equilibrium implies $\Delta^<(p)=-i f_B(p_0) \Delta^-(p)$ and  $\Delta^>(p)=-i(1 + f_B(p_0))\Delta^-(p)$, allowing to interpret the terms in \eqref{GainLoss1} as gain and loss terms with distribution functions $f_B$ and $(1 + f_B)$ for particles in the initial and final states, respectively.
The rate further grows with $h^2$ because the scattering amplitude is proportional to $h$. The result \eqref{Gammasun_small_Meta} exhibits all these features, as one would expect from the general discussion after Eq.~\eqref{Pi_varphi_Def}.

The term $\Gamma_\varphi^{[2]}$, on the other hand, is a non-linear effect that arises due to the time dependence of the quasiparticle properties in the plasma that is induced by their coupling to $\varphi$.  The overall factor $h^2\varphi^2$ in the rate \eqref{GammaTadpole} implies that $\Gamma_\varphi^{[2]}$ can dominate over $\Gamma_\varphi^{[1]}$ for  elongations $\varphi> T$.
The physical reason for this is that the replacement $\phi\to\varphi+\eta$ in the Lagrangian \eqref{Lagrangian} induces a vertex with an effective coupling constant $h\varphi$ that grows with $\varphi$.\footnote{The careful reader might wonder how this can bring powers of $\varphi$ in the numerator, since $\Gamma_\varphi^{[2]}$ arises from the $\varphi$-dependence of the propagators, and therefore one might expect that $h\varphi$ appears in the denominators. Indeed, in \cite{Cheung:2015iqa} the overall factor $\varphi^2$ that enhances this contribution to $\Gamma_\varphi$ for large field elongations was obtained from the small field expansion.
In our current treatment, one power of $\varphi$ is brought into the numerator through the $G_0^{-1}[\varphi]$ in \eqref{Gamma1} when applying \eqref{Gamma1Deriv} to obtain \eqref{Lienland}, and a second power of $\varphi$ in the numerator comes from the $G_0^{-1}[\varphi]$ when applying \eqref{PropExpansion} in \eqref{Torfmoorholm}.
No small field expansion was made to obtain this result.
} A less intuitive property of the result 
\eqref{GammaTadpole} lies in its dependence on $T$. 
The overall factor $1/T$ comes from the distribution function in \eqref{Gdef} and should therefore be interpreted together with the other distribution functions in \eqref{GammaTadpole}. 
Those distribution functions suggest an interpretation in terms of scattering processes in which $\varphi$-quanta of with vanishing momentum and energy $\omega$ are absorbed by $\chi$-quasiparticles with momentum $\p$ and energy $\Omega_\chi$. Since $\Omega_\chi$ (which depends on $\p$, $T$, and $\varphi$) can be larger or smaller than $T$, the quantum statistical factors can either enhance or suppress the integrand \eqref{GammaTadpole}. 
Additional dependencies on $T$ and $\varphi$ are hidden in the $\Omega_\chi$ and $\Gamma_\chi$ in the denominator.
In combination, this can lead to non-trivial parametric dependencies that are discussed in~\cite{Buldgen:2020umy},
including regimes where $\Gamma_\varphi^{[2]}$ decreases with $T$, which naively appears to contradict the microphysical interpretation in terms of scatterings (which should become more frequent as $T$ increases).
A possible explanation for this somewhat unexpected behaviour can be obtained by noticing that the leading processes (e.g.~ $\varphi\varphi\chi\to\chi$) are kinematically forbidden for on-shell quasiparticles, but become allowed for $\omega < \Gamma_\chi$ due to the finite width.
This explains the $1/\Gamma_\chi$ in terms of a near on-shell $\chi$-propagator.\footnote{
Formally the expression \eqref{GammaTadpole} diverges in the limit $\Gamma_\chi\to0$, but we should keep in mind that we implicitly assumed $\Gamma_\chi\gg\omega$ when using the Breit-Wigner approximation, so that this limit should not be taken.
Physically this simply amounts to the assumption that $\tau_{\rm eq}\sim 1/\Gamma_\chi$ is much shorter than the time scale on which $\varphi$ changes.}  
Since $\Gamma_\chi$ typically grows with $T$, this can lead to the counter-intuitive behaviour that $\Gamma_\varphi^{[2]}$ decreases with $T$ in the regime $T>\varphi$.\footnote{
Note also that the $1/\Gamma_\chi$ behaviour in (5.41) is reminiscent of the parametric dependence that one observes for viscosities in the hydrodynamic limit \cite{Hosoya:1983id} (c.f.~also Sec.II.B in~\cite{Jeon:1994if}). In that context an intuitive interpretation is that transport is primarily driven by an exchange of particles between different regions (rather than scatterings between particles), and that this process becomes more efficient when the mean free path in the medium is longer.
}
In any case, the appearance of $\Gamma_\chi$ as a regulator illustrates the necessity to use resummed propagators, which is automatically done the 2PI effective action.

Let us briefly compare the results obtained here to what was found for the same model in \cite{Cheung:2015iqa}. The main difference is that the approach used here does not rely on a small field expansion around $\varphi=0$, i.e., it can be applied to large field elongations, provided that the expansion in $\dot{\varphi}$ holds, and that higher time derivatives of $\varphi$ are negligible.
Further, the method can be generalised to systematically include higher order terms in the expansion in $\dot{\varphi}$. That is, we can treat systems in which the field elongation is large and the field velocity is not-too-small, but the acceleration must remain under control. 
Finally, the method presented here uses a more consistent scheme to include feedback effects of the $\varphi$-evolution on the bath. In \cite{Cheung:2015iqa} the authors made an attempt to extend the range of validity of the effective equation of motion for $\varphi$ by plugging resummed propagators into the loops by hand. This yields the danger of double-counting contributions, and a diagram was forgotten in $\Gamma_\chi$.

\section{Discussion and Conclusion}\label{sec:Conclusion}
In section \ref{sec:markovian} we  presented a model-independent method to derive an effective Markovian equation of the form \eqref{Master} for a slowly evolving scalar field in a time-dependent background.
The effective potential and damping coefficients can be computed from equations \eqref{EffPot} and \eqref{Gamma_functional_taylor_exp}. 
Similar to several approaches previously presented in the literature, our method relies on a separation between the time scale $\tau_{\rm int}$ associated with microscopic interactions and the time scale on which bulk quantities evolve. 
The precise definition of $\tau_{\rm int}$ depends on the physical system under consideration and is discussed in appendix \ref{appendix}. 

The main benefit of the approach presented here lies in its generality.
The method is model-independent and only relies on the separation of scales  $\tau_{\rm int}\flatfrac{\dot{\varphi}}{\varphi}\ll1$.
In particular, it makes no assumption about the spin of the fields that $\phi$ couples to, nor on the type  
of their interactions, and no quasiparticle approximation is needed. Moreover, it is not necessary to specify a particular contour in the complex time plane along which the effective action is defined.
Finally, the master equations 
\eqref{Master}-\eqref{Gamma_functional_taylor_exp}
 in principle permit a computation of $\Gamma$ and $\V$ to all orders in $\dot{\varphi}$, though the computation of terms with $n>1$ can be difficult in practice, depending on the physical problem under consideration. 
 
  For the purpose of the microphysical interpretation in section \ref{sec:LeadingTerm} and the explicit computations in a specific model in section \ref{sec:model} we adopted additional assumptions commonly made in quantum kinetic theory, including 
  the use of the closed contour shown in figure  \ref{Contour-CTP},  a perturbative expansion of the effective action, a near equilibrium assumption for the background, and a Breit-Wigner approximation for the thermal propagators. We emphasise that none of these additional assumptions are necessary for the general method presented in section \ref{sec:markovian}.

Its generality makes the method potentially suitable for the description of a wide range of adiabatic nonequilibrium quantum systems. We primarily have in mind cosmological applications, such as the evolution of the inflaton field during and after cosmic inflation, warm inflation scenarios,  or moduli fields. 
However, in principle it can be applied to any system that can effectively be described by scalar fields, including systems in condensed matter or atomic physics.

\section*{Acknowledgements}
We would like to thank Bj\"orn Garbrecht, Drazen Glavan, Jong Chol Kim, 
Philipp Klose, 
Michael Ramsey-Musolf 
and Oleg Sushkov for inspiring discussions during the work on this project. 
GB acknowledges the support of the National Fund for Scientific Research (F.R.S.- FNRS Beligum) through a FRIA grant. 
MaD would like to thank the Max Planck Institute for Physics (Werner Heisenberg Institut) for their hospitality during the final phase of this work.

\begin{appendix}
\section{Range of validity}\label{appendix}
An important question for practical applications of the method presented here is the range of validity of the approximations that enter our derivation. The two key assumptions in \eqref{RangeOfValidity} are that
the acceleration $\ddot{\varphi}$ and higher derivatives are small enough to be neglected in the expansion \eqref{AD-approx}, 
and that the suppression of the integral kernels in \eqref{FunctionalTaylor}
for $\tau_{\rm int}\flatfrac{\dot{\varphi}}{\varphi}\ll1$ is sufficiently strong to
 apply the approximation \eqref{AD-approx} inside the integrals. 
 The observation that justifies the second assumption is that  $\delta\varphio(x_i) =  \dot{\varphi}(t)(x_i^0-t)$ in \eqref{FunctionalTaylor} 
is always convolved 
with other functions that can be expressed as products of correlators, which are suppressed or oscillate rapidly for separations of the time arguments that are much larger than $\tau_{\rm int}$. 
 We shall first use the explicit results in  section \ref{sec:LeadingTerm} to verify that the expansion parameter that controls the validity of our method is given by $\tau_{\rm int}\flatfrac{\dot{\varphi}}{\varphi}\ll1$.
The expression in \eqref{Pivarphi1insec2} originates from the explicit field dependence of the 2PI effective action when field insertions are attached to explicit vertices in the second diagram of Fig.~\ref{three-loop-diagrams}.
The kernel with which $\delta\varphio(x_i)$ is multiplied is then given by a product of three propagators and ensures a suppression for time separations that exceed $\tau_{\rm int}$.
In \eqref{Pi_varphi_Def} this suppression is guaranteed by the second factor, which contains two propagators, as can be seen in \eqref{from_local_to_non_local}. 
Formally one can express the range of validity of our method as 
\begin{eqnarray}\label{RangeOfValidity}
\tau_{\rm int}^2\ddot{\varphi}/\varphi\ll (\tau_{\rm int} \dot{\varphi}/\varphi)^n
\ll 1,
\end{eqnarray}
where the second inequality is essential to obtain a Markovian equation of motion for $\phi$ in our approach, and the first inequality determines the power $n$ of $\dot{\varphi}$ that can be included consistently without violating the assumption that corrections to the kinetic term $\ddot{\varphi}$ are negligible.
If we want to use near-equilibrium propagators in the computation of the $\Gamma_{\varphi}^{(n)}$ and $\V_\varphi$ we need to impose the additional condition that the evolution of $\varphi$ is slow compared to the relaxation times of all particles that it couples to ($ \dot{\varphi}/\varphi \ll \Gamma_\chi$ for the explicit example in section \ref{sec:model}). While this last condition is important if one aims for semi-analytic results, it is strictly not necessary to obtain a Markovian equation. 

Turning these considerations into a more quantitative statement requires a specification of the time scale $\tau_{\rm int}$, or at least its parametric dependence. 
For the following discussion we adopt the symbolic notation $\omega \sim \dot{\varphi}/\varphi$. Assuming for simplicity that $\ddot{\varphi}/\varphi \sim \omega^2$ the condition for the applicability of our method  reads 
\begin{eqnarray}\label{ValidityCondition}
\tau_{\rm int} \omega \ll 1.
\end{eqnarray}
An exponential suppression 
of the integral kernels in \eqref{FunctionalTaylor}
that allows for a clean definition of $\tau_{\rm int}$ only kicks in for separations that are comparable to the typical damping/relaxation time in the medium in which $\phi$ resides. Using the notation of section \ref{sec:model} this would amount to $\tau_{\rm int} \sim 1/\Gamma_\chi$. 
We first consider weakly damped systems with $\omega \gg \Gamma_\varphi$. 
If one, for instance, studies small deviations around the minimum of $\V$, one obtains $\partial_\varphi\V\sim M_\phi^2\varphi$. 
For weak damping 
$\Gamma_\varphi\ll M_\phi$ this amounts to $\omega \sim M_\phi$, 
so condition \eqref{ValidityCondition} would read $M_\phi \ll \Gamma_\chi$. This condition can easily be satisfied if $\phi$ is placed in a strongly interacting thermal bath with temperature $T\gg M_\phi$. 
In the toy model considered in section \eqref{sec:model} with $\mathcal{L}_{\chi {\rm int}} = \lambda_\chi \chi^4/4!$, it can be achieved if $T\gg \varphi , M_\phi$ and $\lambda_\chi \gg h$. 
For strong damping one may take $\omega\sim \Gamma_\varphi$, in which case condition \eqref{ValidityCondition} reads $\Gamma_\varphi \ll \Gamma_\chi$, which simply amounts to $\lambda_\chi \gg h$.
This situation is, for instance, encountered frequently in the cosmological study of hidden particles \cite{Agrawal:2021dbo} or during warm inflation.
In both cases we find that the condition reads $\omega/\Gamma_\chi \ll1$ (hence $\tau_{\rm int}\sim 1/\Gamma_\chi$), but this translates into different constraints on the model parameters because $\omega$ is either identified with $M_\phi$ or $\Gamma_\varphi$.

The range of validity of our method is in fact larger than these estimates suggest because the integral kernels in \eqref{FunctionalTaylor} can  effectively be suppressed on time scales $\tau_{\rm int} < 1/\Gamma_\chi$. To see this, we consider the equation
\begin{eqnarray}
\ddot{\varphi}(t) + \omega^2\varphi(t) + \int dt' \Uppi(t-t')\varphi(t') = 0,
\end{eqnarray}
which one would obtain from a Taylor expansion of \eqref{FunctionalTaylor} in field space around $\bar{\varphi}=0$, cf.~e.g.~\cite{Cheung:2015iqa}. 
Following our approach, we can introduce $z=t - t'$ and expand
\begin{eqnarray}\label{derivativesexpansion}
\int dz \Uppi(z)\varphi(t-z) = \sum_k \frac{\partial^k\varphi}{\partial t^k}\frac{1}{k!}\int dz \ (-z)^k \ \Uppi(z).
\end{eqnarray}
The applicability of our method now relies on the assumption that the first term in this series dominates. 
We know for sure that $\Uppi(z)$ is exponentially suppressed for $|z|>\Gamma_\chi$, as all thermal correlation functions are subject to this suppression \cite{Garbrecht:2011xw}. Using a mock kernel 
$\Uppi(z)\propto h^2 \omega^3e^{-|z|\Gamma_\chi}$, 
explicit computation shows that each integral in the series \eqref{derivativesexpansion} is convergent, 
and that the effective expansion parameter of the series is $\omega/\Gamma_\chi$,  
which is consistent with the assumption that the relaxation time scale $1/\Gamma_\chi$ within the bath is much shorter than the time scale $1/\omega$ on which $\varphi$ evolves.
Hence, we confirm that the choice $\tau_{\rm int} = 1/\Gamma_\chi$ leads to a well-defined expansion scheme.

We now proceed to investigate whether choices $\tau_{\rm int} < \Gamma_\chi$ still give a well-defined expansion scheme. 
In addition to the exponential suppression $\Uppi(z)\propto e^{-|z|\Gamma_\chi}$, $\Uppi(z)$ typically exhibits a power law suppression and oscillatory behaviour on shorter time scales. This behaviour has been studied in detail in \cite{Gautier:2012vh} for a $g\phi\chi^2$ interaction and it was found that $\Uppi(z) \propto g^2 m_\chi\frac{\cos(m_\chi z + \pi/4)}{(m_\chi z)^{3/2}}(f_B(m_\chi) + \frac{1}{2})$, where the exponential suppression with $\Gamma_\chi$ that kicks in for larger separation has been neglected. 
For illustrative purposes we simply multiply this expression by $e^{-\Gamma_\chi|z|}$ to include this suppression. 
Plugging this into \eqref{derivativesexpansion} we have to solve integrals of the kind
\begin{eqnarray}
 \frac{\partial^k\varphi}{\partial t^k}
 \frac{g^2 m_\chi}{k!} \int dz (-z)^k\frac{\cos(m_\chi z + \pi/4)}{(m_\chi z)^{3/2}}\left(f_B(m_\chi) + \frac{1}{2}\right)e^{-\Gamma_\chi|z|}.
\end{eqnarray}
This integral can be solved analytically,
\begin{eqnarray}
\lefteqn{\frac{(-1)^kg^2}{\sqrt{2}}\left(\frac{\omega}{m_\chi}\right)^k \left(1+\frac{\Gamma_\chi^2}{m_\chi^2}\right)^{(1-2k)/4}\frac{\Gamma_E (k-1/2)}{k!
} \left(f_B(m_\chi) + \frac{1}{2}\right)}\nonumber\\
&\times&\bigg[\cos\left (\frac{1}{2}(1-2k) \arctan \left(\frac{m_\chi}{\Gamma_\chi}\right)\right)
            \ + \ \sin\left (\frac{1}{2}(1-2k) \arctan \left(\frac{m_\chi}{\Gamma_\chi}\right)\right)\bigg],
\end{eqnarray}
where we have replaced $\frac{d^k\varphi}{d t^k} \to \omega^k$ and $\Gamma_E$ is the Euler Gamma function.
Expanding to leading order in $\Gamma_\chi/m_\chi$, we find that the $k$-th term in the series \eqref{derivativesexpansion} reads

\begin{equation}
     (-1)^kg^2 \left(\frac{\omega}{m_\chi}\right)^k
     \frac{\Gamma_E
                (k-1/2)}{k!} \bigg[\cos\left (\frac{k \pi}{2}\right)
     +\frac{\Gamma_\chi}{m_\chi}\,\left(k-\frac{1}{2}
     \right)\,\sin\left (\frac{k \pi}{2}\right)\bigg]
     \left(f_B(m_\chi) + \frac{1}{2}\right).
     \end{equation}
These expressions indicate that the expansion parameter that controls the convergence of the series is in fact $\omega/m_\chi$. Hence, we can identify $\tau_{\rm int}\sim 1/m_\chi$
in \eqref{ValidityCondition}. This is a considerably weaker criterion than $\tau_{\rm int}\sim 1/\Gamma_\chi$, and it would become even weaker if one were to consider screening effects that dress $m_\chi$ into a (typically larger) effective in-medium mass $M_\chi$.

Of course, the derivation presented here relies on the functional form of $\Uppi(z)$ that we used, which is specific to the particular interaction considered in \cite{Gautier:2012vh}. 
However, there are reasons to believe that the behaviour observed here (and in particular the appearance of the ratio between $\omega$ and an effective particle mass $M_\chi$ as an expansion parameter) is representative for a wider class of similar systems,
as the behaviour of loop integrals is primarily determined by the frequencies that appear in the loop, with the width giving a small correction that typically only matters near thresholds.
This suggests that 
an effective suppression of memory on time scales shorter than $1/\Gamma_\chi$ can occur in at least some physical systems, in which one can relate $\tau_{\rm int}$ to the frequencies in the bath  (i.e.,  $\tau_{\rm int}\sim \#/M_\chi$ or more generally $\tau_{\rm int}\sim \#/\Omega_\chi$, with $\#$ some numerical factor).
On the other hand, the study in \cite{Berges:2005md}
indicated that $\tau_{\rm int}$ should be identified with the relaxation time in the bath ($\tau_{\rm int}\sim 1/\Gamma_\chi$), based on a comparison between the numerical solutions for the full non-local equations for the two-point functions $\Delta^\pm$ and Markovianised approximations. 
One difference is that the authors of \cite{Berges:2005md} considered initial conditions far from thermal equilibrium, while our simple example assumes a setup in which only one degree of freedom is out of equilibrium and couples to a large thermal bath. 
Another difference is that the authors of \cite{Berges:2005md} 
investigated the relaxation of the particle occupation numbers characterised by the statistical propagator $\Delta^+$ (and assumed $\varphi=0$), while we study the time evolution of $\varphi$. 
Finally, they considered a system with only one single (comparably strong) coupling constant, while our example involves a hierarchy between the interaction strength of $\phi$ with the bath and the couplings of the bath's constituents amongst each other. 
A detailed investigation of all these aspects goes beyond the scope of the this work. 
In the present context we take the differences as an indicator 
that the smallest choice of $\tau_{\rm int}$ that can be justified depends on the system under investigation. 
The choice $\tau_{\rm int}=1/\Gamma_\chi$ is a safe one, but potentially too conservative, while the choice  $\tau_{\rm int}\sim \#/\Omega_\chi$ may only be justified under specific conditions. 
Finally, we should add that the difficulties in the identification of the time scale associated with memory loss 
are not specific to our method,
and similar ambiguities exist in other approaches, 
which provided one of the main motivations for the numerical study in \cite{Berges:2005md}.

\end{appendix}

\bibliographystyle{JHEP}
\bibliography{main}

\end{document}